\documentclass[preprint,superscriptaddress,amsmath,amssymb,aps,physrev,floatfix,longbibliography,showkeys]{revtex4-2}
\usepackage[utf8]{inputenc}
\usepackage{graphicx}
\usepackage[table]{xcolor}
\usepackage{tikz}
\usepackage{url}
\usepackage{adjustbox}
\usetikzlibrary{automata,positioning,calc,arrows.meta}
\usepackage{xr-hyper}
\usepackage{booktabs}
\usepackage[hidelinks]{hyperref}
\IfFileExists{supplementary.aux}{%
  \externaldocument[supp-][nocite]{supplementary}[supplementary.pdf]%
}{%
  \externaldocument[supp-][nocite]{crossrefs-supplementary}[supplementary.pdf]%
}

\makeatletter
\let\originalFrontmatterAbstract\frontmatter@abstract@produce
\renewcommand{\frontmatter@abstract@produce}{%
  \begingroup
  \@booleanfalse\preprintsty@sw
  \originalFrontmatterAbstract
  \endgroup
}
\makeatother

\begin{document}
\title{Locating Debunkers for Rumor Mitigation in Directed Networks}

\author{Kasra Ghorbani}
\thanks{These authors contributed equally to this work.}
\affiliation{CompleX Lab, School of Computer Science and Engineering,
University of Electronic Science and Technology of China,
Chengdu 610054, People's Republic of China}
\affiliation{College of Engineering, University of Tehran, Tehran, Iran}

\author{Xinshan Jiao}
\thanks{These authors contributed equally to this work.}
\affiliation{CompleX Lab, School of Computer Science and Engineering,
University of Electronic Science and Technology of China,
Chengdu 610054, People's Republic of China}

\author{Tao Zhou}
\email[Contact author: ]{zhutou@ustc.edu}
\affiliation{CompleX Lab, School of Computer Science and Engineering,
University of Electronic Science and Technology of China,
Chengdu 610054, People's Republic of China}

\begin{abstract}
Rumors can spread widely through social networks, making the selection of a limited number of debunkers to disseminate corrective information important. Existing centrality-based methods only consider the individual influence of a node on spreading processes. Those individually influential nodes often have heavily overlapping influence ranges, resulting in a limited number of users they can cover. To address this problem, we propose the so-called Hybrid Marginal Influence (HMI) framework, which combines node centrality with additional audience coverage provided by each candidate and adaptively balances these two factors through a greedy selection procedure. We further develop the Debunker-Augmented SIR (DA-SIR) model for the propagation of rumors and corrective information. Experiments on ten real-world directed networks show that HMI generally selects debunkers with broader audience coverage than conventional centrality-based methods, accompanied by a lower final cumulative fraction of users who repost the rumor.
\end{abstract}

\maketitle

\section{Introduction}

Rumors can spread rapidly through social networks, reaching large audiences through repeated sharing and reposting \cite{vosoughi2018spread}. Disseminating corrective information is an effective way to suppress their spread \cite{ecker2022psychological,budak2011limiting}, but the effectiveness of this intervention depends on whether corrective information can reach sufficient users exposed to the rumor. In practice, only a fraction of users may be available to actively disseminate corrective information. This creates a practical challenge: when the intervention budget is limited, which users should be selected as debunkers? The answer depends not only on the ability of individual users to spread information but also on the collective reach of the selected group.

A natural strategy is to recruit users who occupy important positions in the network. Centrality measures provide a convenient way to identify such users, as they characterize different aspects of structural importance \cite{lu2016vitalnodes,bi2026performances}. Well-known examples include Degree \cite{freeman1978centrality}, PageRank \cite{page1999pagerank}, Katz \cite{katz1953status}, Betweenness \cite{freeman1977betweenness}, the k-core (k-shell) index \cite{kitsak2010identification}, and the network H-index \cite{lu2016hindex}. However, selecting individually important users does not necessarily produce an effective group of users \cite{zhang2016voterank,ramrakhiyani2024influence}. Highly central users may reach many of the same users, meaning that their corrective information repeatedly reaches similar audiences. Although repeated exposure may still be useful \cite{centola2010spread,lu2011smallworld}, each additional debunker contributes less new audience coverage. In contrast, a debunker with lower centrality may reach a different part of the network and introduce corrective information to audiences that would otherwise remain uncovered.

Simply maximizing audience coverage, however, may also be insufficient. Users reaching relatively few or peripheral audience members may contribute new coverage but lack structural importance. Conversely, highly central users may remain valuable even when some of their audiences have already been reached. Effective selection therefore requires considering two complementary properties: the structural importance of each candidate and the additional audience coverage that the candidate contributes to the selected group. The key question is how to balance these properties so that a limited number of debunkers can retain the advantages of individually influential nodes while reducing redundant audience coverage. This problem is closely related to influence maximization and network-based rumor control. Influence maximization studies how to select a set of seed nodes to achieve a large collective spreading effect \cite{kempe2003maximizing,chen2009efficient}. Previous studies on rumor mitigation have examined information blocking and the dissemination of corrective information through selected users \cite{budak2011limiting,he2012influence,xie2023minimizing,li2025efficient}.
Other studies have developed methods for identifying the influential node set. Zhang \textit{et al.}~\cite{zhang2016voterank} reduce the voting ability of neighbors of selected nodes to encourage more dispersed selections. Morone and Makse~\cite{morone2015influence} demonstrate the importance of considering collective influence. Fan \textit{et al.}~\cite{fan2021characterizing} identify influential nodes scattered across different regions by counting the frequency of a node's participation in the shortest cycles. Ramrakhiyani \textit{et al.}~\cite{ramrakhiyani2024influence} use network partitioning to distribute influential spreaders across different regions.

Motivated by the above perspective, we propose the Hybrid Marginal Influence (HMI) framework, which combines node centrality with marginal audience coverage under a limited selection budget. HMI evaluates each candidate according to both its structural importance and its contribution additional to that of the selected debunkers. Rather than treating all reachable users equally, the framework assigns decreasing weights to users who have already been covered by multiple selected debunkers. It then adaptively balances the centrality and marginal coverage through a greedy procedure. Because HMI can incorporate different centrality measures, it provides a general approach to improve centrality-based debunker-selection strategies.

To evaluate the effectiveness of different debunker-selection strategies, we propose the Debunker-Augmented SIR (DA-SIR) model, motivated by the classical SIR model \cite{kermack1927sir,pastor2015epidemic} and its variants for rumor spreading \cite{zhao2013sir,fan2013rumor,granell2013dynamical,wang2017dynamics}. The DA-SIR model describes rumor spreading and corrective-information dissemination in directed social networks by introducing two types of debunkers: Selected Debunkers (SD), who are recruited to initiate corrective information, and Temporary Debunkers (TD), who propagate corrective information after receiving it. We evaluate four HMI variants against eight baseline methods on ten directed networks. HMI achieves greater direct audience coverage in 38 of 40 comparisons with its corresponding centrality methods and reduces the fraction of ever-infected nodes in 37 of 40 comparisons.

\section{Modeling Rumor Spreading with Debunkers}
\label{sec:model}

We denote by \(G=(V,E)\) a directed graph, where \(V\) is the set of nodes and \(E\subseteq V\times V\) is the set of directed exposure relations. An edge \((u,v)\in E\) means that node \(u\) can receive information from node \(v\). Information therefore propagates from \(v\) to \(u\), opposite to the edge direction. We apply this convention to all networks considered in the experiments, treating their directed relations as potential information-transmission pathways. On this network, we consider rumor spreading and corrective-information dissemination as two competing processes \cite{wang2019coevolution}. To describe how corrective information can prevent or interrupt rumor propagation, we extend the classical Susceptible--Infected--Recovered (SIR) model \cite{kermack1927sir,pastor2015epidemic} by introducing two debunker states, yielding the so-called Debunker-Augmented SIR (DA-SIR) model.

The DA-SIR model contains five states. Susceptible nodes (\(S\)) have not propagated either type of information. Infected nodes (\(I\)) have begun propagating the rumor, while recovered nodes (\(R\)) have stopped or avoided propagating the rumor. Two additional states describe the propagation of corrective information. Selected Debunkers (\(SD\)) are nodes chosen to initiate corrective-information spreading at a specified intervention time. Before activation, they remain in \(SD\), do not transmit information, and cannot become infected. Temporary Debunkers (\(TD\)) are nodes that begin propagating corrective information with a certain probability after receiving it from an active debunker. Both \(SD\) and \(TD\) enter \(R\) when their active corrective-information spreading period ends.

A susceptible node may enter \(I\) and begin propagating the rumor, enter \(TD\) and begin propagating corrective information, or move directly to \(R\) without propagating either type of information. An infected node may also enter \(TD\) after receiving corrective information, which terminates its ongoing rumor spreading. A recovered node may still enter \(TD\) if it has not previously propagated corrective information. Each node can initiate the propagation of the rumor at most once and the propagation of corrective information at most once.

Exposure-driven responses are controlled by three parameters. The parameter \(\beta\) is the probability that a susceptible node begins propagating the rumor upon one exposure to an infected node. The parameter \(\gamma\) is the probability that a node in \(S\), \(I\), or \(R\) begins propagating corrective information upon one exposure to a debunker. The transition from \(R\) to \(TD\) is permitted only for nodes that have not previously propagated corrective information. The parameter \(\alpha\) is the probability that a susceptible node moves directly to \(R\) upon one exposure to a debunker.

\begin{figure*}[t]
\centering
\includegraphics[width=\textwidth]{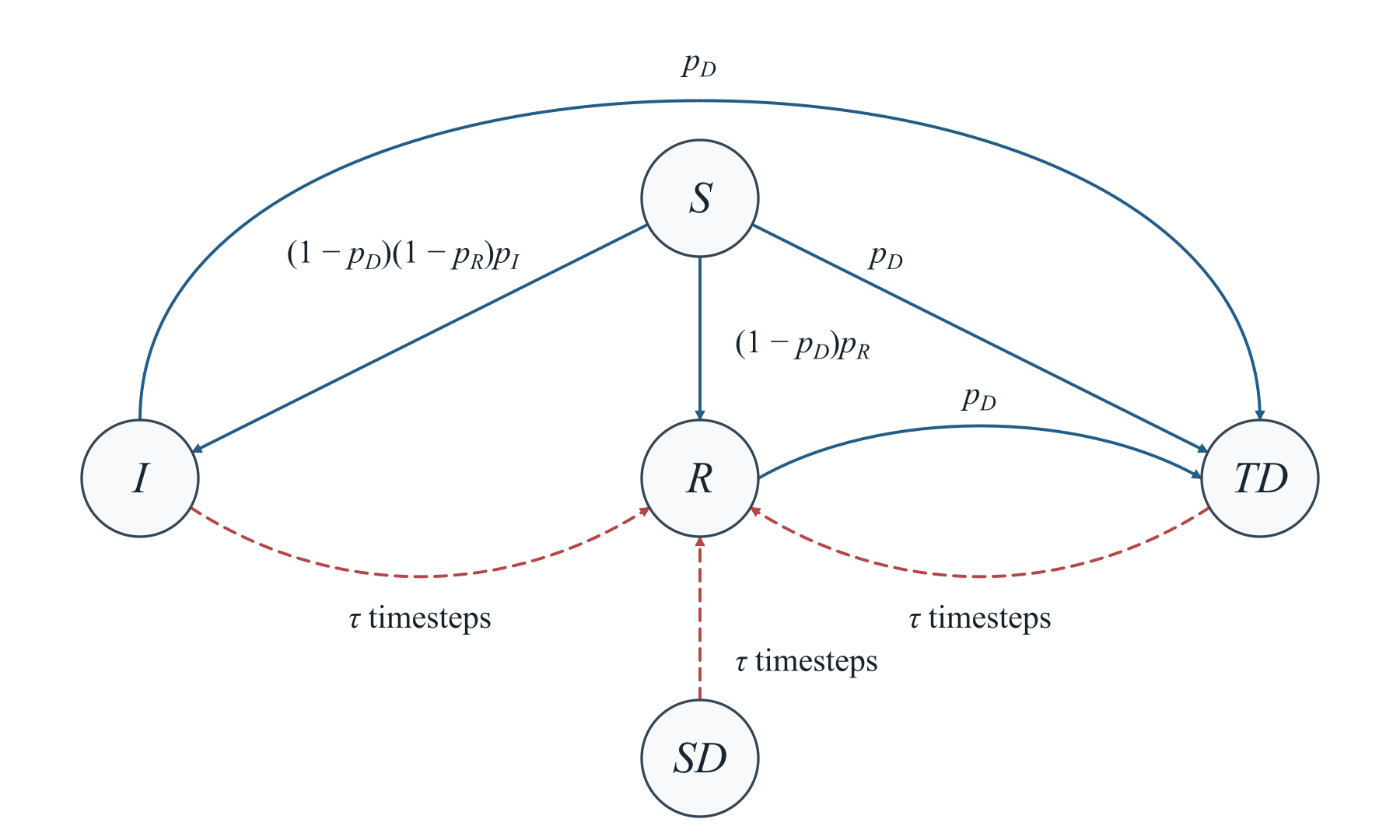}
\caption{State transitions in the DA-SIR model. Solid blue arrows represent exposure-driven transitions with probabilities labeled next to the arrows. The priority of each transition is described in the main text. Dashed red arrows represent transitions caused by the end of an active spreading period lasting \(\tau\) timesteps. Such transitions occur only when no higher-priority exposure-driven transition succeeds.}
\label{fig:state-diagram}
\end{figure*}

The positive integer \(\tau\) specifies the duration of active information spreading. Once a node begins propagating the rumor, it can transmit the rumor for at most \(\tau\) timesteps (it may enter \(TD\) before \(\tau\) timesteps). A node that begins propagating corrective information can transmit it for \(\tau\) timesteps after activation. Updates are synchronous. At each timestep, active transmitters make independent exposure attempts for each applicable response type, and these attempts may recur during the active spreading period. All exposure attempts are evaluated before state changes take effect, and each node undergoes at most one state transition per timestep. The intervention time and initialization procedure are described in the experimental setup at the beginning of Section~\ref{sec:analysis}.

Let \(k_I\) and \(k_D\) denote the number of infected nodes and debunkers, respectively, from which the target node can receive information. Under independent exposure attempts, the probabilities of at least one successful rumor-propagation, corrective-information-propagation, and direct-recovery response in one timestep are
\begin{align}
p_I(k_I)&=1-(1-\beta)^{k_I},\\
p_D(k_D)&=1-(1-\gamma)^{k_D},\\
p_R(k_D)&=1-(1-\alpha)^{k_D}.
\end{align}
These are response-attempt probabilities rather than final state-transition probabilities. When multiple responses are possible, corrective-information propagation takes precedence over direct recovery, which in turn takes precedence over rumor propagation. For a susceptible node, the resulting state-transition probabilities are
\begin{align}
\Pr(S\!\to\!TD)&=p_D,\\
\Pr(S\!\to\!R)&=(1-p_D)p_R,\\
\Pr(S\!\to\!I)&=(1-p_D)(1-p_R)p_I,\\
\Pr(S\!\to\!S)&=(1-p_D)(1-p_R)(1-p_I).
\end{align}
Here, \(p_I\), \(p_D\), and \(p_R\) are calculated at the corresponding \(k_I\) and \(k_D\).

An infected node can enter \(TD\) with probability \(p_D\) at each timestep (\(p_D\) can change over time); otherwise, it enters \(R\) when its active rumor-spreading period ends. A recovered node can enter \(TD\) only if it has not previously propagated corrective information. Temporary and selected debunkers enter \(R\) when their active corrective-information spreading period ends. A duration-based transition takes effect only when no higher-priority exposure-driven transition succeeds. Figure~\ref{fig:state-diagram} summarizes the transitions in the DA-SIR model.

\begin{figure*}[t]
\centering
\begin{tikzpicture}[
    >=Latex,
    font=\small,
    node distance=1.5cm,
    base/.style={circle, draw=black!55, fill=gray!3, thick, minimum size=6mm, inner sep=0pt},
    reached/.style={circle, draw=teal!55!black, fill=teal!18, thick, minimum size=6mm, inner sep=0pt},
    selected/.style={circle, draw=black, fill=gray!75, text=white, thick, font=\footnotesize\bfseries, minimum size=6.5mm, inner sep=0pt},
    candA/.style={circle, draw=violet!80!black, fill=violet!12, thick, font=\small\bfseries, minimum size=7mm, inner sep=0pt},
    candB/.style={circle, draw=orange!90!black, fill=orange!18, thick, font=\small\bfseries, minimum size=7mm, inner sep=0pt},
    edge_old/.style={draw=teal!45!black, thick},
    edge_overlapA/.style={draw=violet!65, thick, dashed},
    edge_newA/.style={draw=violet!85!black, thick},
    edge_overlapB/.style={draw=orange!70!black, thick, dashed},
    edge_newB/.style={draw=orange!90!black, thick}
]

\node[selected] (d1) at (1.4, 5.0) {$d_1$};
\node[selected] (d2) at (1.4, 2.5) {$d_2$};

\node[reached] (r1) at (3.4, 6.0) {};
\node[reached] (r2) at (4.0, 5.0) {};
\node[reached] (r3) at (3.7, 3.8) {};
\node[base] (r4) at (4.0, 2.5) {};
\node[reached] (r5) at (3.4, 1.5) {};

\node[candA] (va) at (6.3, 5.0) {$v_A$};
\node[base] (a1) at (8.3, 5.7) {};
\node[base] (a2) at (8.3, 4.3) {};

\node[candB] (vb) at (6.3, 2.5) {$v_B$};
\node[base] (b1) at (8.3, 3.5) {};
\node[base] (b2) at (8.5, 2.5) {};
\node[base] (b3) at (8.3, 1.5) {};

\draw[edge_old, ->] (r1) -- (d1);
\draw[edge_old, ->] (r2) -- (d1);
\draw[edge_old, ->] (r3) -- (d1);

\draw[edge_old, ->] (r3) -- (d2);
\draw[edge_old, ->] (r5) -- (d2);

\draw[draw=black!55, thick, ->] (r4) -- (r3);
\draw[draw=black!55, thick, ->] (r4) -- (r5);

\draw[edge_overlapA, ->] (r1) -- (va);
\draw[edge_overlapA, ->] (r2) -- (va);
\draw[edge_overlapA, ->] (r3) -- (va);

\draw[edge_newA, ->] (a1) -- (va);
\draw[edge_newA, ->] (a2) -- (va);

\draw[edge_newB, ->] (r4) -- (vb);

\draw[edge_newB, ->] (b1) -- (vb);
\draw[edge_newB, ->] (b2) -- (vb);
\draw[edge_newB, ->] (b3) -- (vb);

\node[align=left, font=\footnotesize, text=violet!80!black, anchor=west] at (9.0, 5.0) {
    Centrality: higher\\
    Marginal gain: lower
};

\node[align=left, font=\footnotesize, text=orange!85!black, anchor=west] at (9.0, 2.5) {
    Centrality: lower\\
    Marginal gain: higher
};

\begin{scope}[shift={(0, 7.25)}]
    \node[selected, minimum size=5mm] (n1) at (0,0) {};
    \node[anchor=west, xshift=1.5mm, font=\footnotesize]
        (legendSelected) at (n1.east) {Selected debunker};

    \node[reached, minimum size=5mm, anchor=west, xshift=3mm]
        (n2) at (legendSelected.east) {};
    \node[anchor=west, xshift=1.5mm, font=\footnotesize]
        (legendCovered) at (n2.east) {Already covered};

    \node[base, minimum size=5mm, anchor=west, xshift=3mm]
        (n3) at (legendCovered.east) {};
    \node[anchor=west, xshift=1.5mm, font=\footnotesize]
        (legendUncovered) at (n3.east) {Not covered};

    \coordinate (overlapStart) at ([xshift=3mm]legendUncovered.east);
    \draw[draw=black!55, thick, dashed]
        (overlapStart) -- ++(5mm,0) coordinate (overlapEnd);
    \node[anchor=west, xshift=1.5mm, font=\footnotesize]
        (legendOverlap) at (overlapEnd) {Overlap};

    \coordinate (newStart) at ([xshift=3mm]legendOverlap.east);
    \draw[draw=black!80, thick]
        (newStart) -- ++(5mm,0) coordinate (newEnd);
    \node[anchor=west, xshift=1.5mm, font=\footnotesize]
        at (newEnd) {New coverage};
\end{scope}

\end{tikzpicture}

\caption{
Illustration of the HMI scoring logic. This example considers degree as the base centrality. Directed edges follow the convention that \(u\rightarrow v\) means that \(u\) can receive information from \(v\). Candidate \(v_A\) has higher centrality but more audience overlap with the selected debunkers, whereas \(v_B\) has lower centrality but greater marginal coverage gain.
}
\label{fig:hmi_single_network}
\end{figure*}
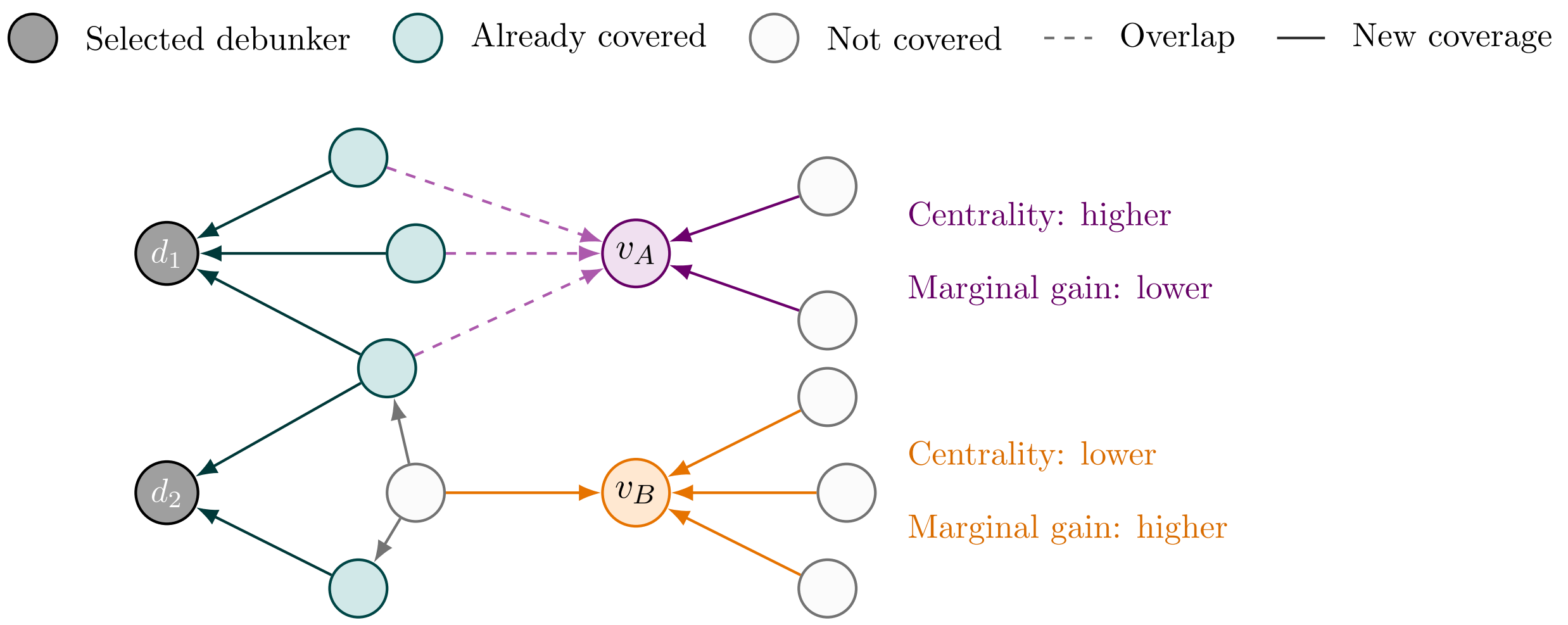

\section{HMI Framework}
\label{sec:hmi}

The Hybrid Marginal Influence (HMI) framework selects debunkers by considering both the importance of individual nodes and the additional audience coverage they bring to the selected set. That is to say, a candidate is favored not only when it has high centrality, but also when it reaches nodes that have difficulty receiving corrective information from the current debunker set. Figure~\ref{fig:hmi_single_network} illustrates this idea.

Let \(K\) (\(1\leq K\leq |V|\)) denote the number of debunkers to select. We use \(D\subseteq V\) to represent the set of selected debunkers, and the coverage neighborhood of a node \(v\in V\) is defined as
\begin{equation}
\mathcal{N}(v)
=
\{v\}\cup\{u\in V:(u,v)\in E\}.
\end{equation}
This neighborhood contains \(v\) itself and all nodes that can directly receive information from \(v\). When several debunkers are selected, their coverage neighborhoods may overlap. To measure this overlap, we define \(m_D(u)\) as the number of selected debunkers whose coverage neighborhoods contain node \(u\). For example, \(m_D(u)=0\) means that \(u\) has not been covered by any selected debunker, whereas \(m_D(u)=2\) means that two selected debunkers can directly reach it.

Reaching a previously uncovered node extends the audience of the selected set. Reaching a covered node may still be useful, since an earlier exposure does not necessarily produce a response \cite{sanderson2023listening}, but the additional benefit should decrease as coverage accumulates\cite{centola2010spread,lu2011smallworld}. HMI therefore assigns a weight \(w_m\) to an additional coverage of a node that has already been covered \(m\) times. The weight is largest when \(m=0\) and decreases as \(m\) increases.

To determine \(w_m\), we consider an auxiliary probability \(q\in(0,1)\) that a single corrective exposure produces a relevant response. Suppose responses to different covering debunkers are conditionally independent given \(q\). If a node receives \(m\) exposures, the probability of at least one response is \(1-(1-q)^m\). Adding one more exposure increases this probability by
\begin{equation}
\delta(m,q)
=
\left[1-(1-q)^{m+1}\right]
-
\left[1-(1-q)^m\right]
=
q(1-q)^m.
\end{equation}
This expression provides a simple interpretation of diminishing returns. Each additional exposure can produce a response only if the previous exposures have not already done so. Consequently, its marginal contribution decreases with \(m\).

The response probability \(q\), however, is generally unknown. Rather than choosing an arbitrary fixed value, we average the marginal gain over a probability distribution for \(q\). Specifically, we use the Jeffreys prior for a Bernoulli probability parameter \cite{jeffreys1946invariant}, where \(q\) obeys the probability density function
\begin{equation}
\pi(q)
=
\frac{1}{B(\tfrac12,\tfrac12)}
q^{-1/2}(1-q)^{-1/2}.
\end{equation}
The Jeffreys prior provides a standard, parameterization-invariant choice for the unknown probability. Here, it is used to construct a fixed coverage-weighting rule without estimating a response probability from spreading data. We therefore marginalize over \(q\) under the Jeffreys prior to obtain the marginal weights for repeated coverage, which gives
\begin{equation}
\delta_m
=
\mathbb{E}_{q}\!\left[q(1-q)^m\right]
=
\frac{1}{B(\tfrac12,\tfrac12)}
\int_0^1 q^{1/2}(1-q)^{m-1/2}\,dq
=
\frac{B(\tfrac32,m+\tfrac12)}
{B(\tfrac12,\tfrac12)}.
\end{equation}
The quantity \(\delta_m\) represents the expected marginal response gain associated with one additional coverage after \(m\) previous coverages under the auxiliary probability model. We normalize it by the value for the first coverage, \(\delta_0=\tfrac12\), and define \(w_m=\delta_m/\delta_0\). Using the beta-function identity, the weights can be calculated recursively as
\begin{equation}
w_0=1,
\qquad
w_{m+1}
=
w_m\frac{m+\tfrac12}{m+2},
\label{eq:jeffreys_recurrence}
\end{equation}
and a few initial weights are \(w_0=1\), \(w_1=1/4\), \(w_2=1/8\), and \(w_3=5/64\).
Thus, a previously uncovered node receives full weight, while an additional coverage of a node already covered once receives one-quarter of that weight. The weight continues to decrease with repeated coverage but remains positive.

Using these weights, we define the marginal coverage gain of a candidate \(v\), given the selected set \(D\), as
\begin{equation}
G(v\mid D)
=
\frac{1}{|V|}
\sum_{u\in\mathcal{N}(v)}
w_{m_D(u)}.
\label{eq:jeffreys_gain}
\end{equation}
This expression sums the weighted contributions of all nodes in the candidate's coverage neighborhood. The factor \(1/|V|\) scales the gain by network size. Initially, when \(D=\emptyset\), every node has zero coverage multiplicity and \(w_0=1\), so
\[
G(v\mid\emptyset)
=
\frac{|\mathcal{N}(v)|}{|V|}.
\]
At this stage, the gain is proportional to the size of the candidate's coverage neighborhood. As debunkers are selected, the weights of previously covered nodes decrease. Consequently, a candidate reaching many uncovered nodes produces a large marginal gain.

Marginal coverage alone, however, does not consider the individual importance of a candidate. HMI therefore combines \(G(v\mid D)\) with a base-centrality score \(C(v)\).
Let \(D_t\) denote the debunker set after \(t\) selections, with \(D_0=\emptyset\), \(|D_t|=t\), and the remaining candidate set \(\mathcal{R}_t=V\setminus D_t\). Since centrality and marginal coverage have different numerical ranges, HMI applies min--max normalization to each score over the remaining candidates. For the centrality score \(C(v)\) of node \(v\), at iteration \(t\) we define
\begin{equation}
\widehat{C}_t(v)
=
\frac{C(v)-\min_{u\in\mathcal{R}_t}C(u)}
{\max_{u\in\mathcal{R}_t}C(u)-\min_{u\in\mathcal{R}_t}C(u)}.
\end{equation}
We define the normalized marginal coverage gain \(\widehat{G}_t(v\mid D_t)\) in a similar way. Both normalized scores lie between zero and one. Normalization is repeated after each selection because the remaining candidate set changes and the marginal coverage gains are updated.

The two normalized scores are combined through
\begin{equation}
\mathrm{HMI}_t(v)
=
\theta_t\,\widehat{C}_t(v)
+
(1-\theta_t)\,
\widehat{G}_t(v\mid D_t),
\label{eq:HMI}
\end{equation}
where \(\theta_t\in[0,1]\) controls their relative contributions. Rather than fixing \(\theta_t\) in advance, HMI determines it from the variation of the two normalized scores among the remaining candidates. Motivated by the dispersion-based weighting principle~\cite{diakoulaki1995critic}, we assign greater weight to the score with higher variation across candidates, as it provides greater discrimination among them. Specifically, we set \(\theta_t\) as
\begin{equation}
\theta_t
=
\frac{\sigma_C(t)}
{\sigma_C(t)+\sigma_G(t)},
\end{equation}
where \(\sigma_C(t)\) and \(\sigma_G(t)\) are the standard deviations of \(\widehat{C}_t(v)\) and \(\widehat{G}_t(v\mid D_t)\), respectively, over the remaining candidates \(v\in\mathcal{R}_t\). 

At each iteration \(t\), we select the remaining candidate with the highest \(\mathrm{HMI}_t(v)\) and add it to the selected set. This process continues until \(K\) debunkers have been selected. We construct four variants using Degree, PageRank, Katz, and Betweenness centrality, named HMI-Degree, HMI-PageRank, HMI-Katz, and HMI-Betweenness. They differ only in the base-centrality score \(C(v)\), while the marginal coverage calculation, adaptive weighting, and greedy selection procedure are the same.

\section{Results}
\label{sec:analysis}
\label{sec:experimental_protocol}

We evaluate four HMI variants and eight baseline methods, including Degree, PageRank, Katz, Betweenness, Leverage Centrality (LC), LeaderRank, VoteRank, and Collective Influence (CI), on ten real-world directed networks: Twitter~\cite{mcauley2012ego,snapnets}, Google+~\cite{fire2013link,rossi2015network}, PGP Web of Trust~\cite{richters2011trust}, Facebook Wall Posts~\cite{viswanath2009facebook,kunegis2013konect}, Epinions~\cite{richardson2003trust,snapnets}, Digg Friends~\cite{lerman2016digg,hogg2012social}, Weibo Followers~\cite{weibo2016figshare}, Munmun Twitter~\cite{dechoudhury2010sampling,kunegis2013konect}, YouTube Social Links~\cite{mislove2007measurement,kunegis2013konect}, and Enron Email~\cite{rossi2015network}. For each network, duplicate directed edges are merged, and only the largest weakly connected component is retained. Table~\ref{tab:dataset_statistics_compact} reports the elementary statistics of these networks.

\begin{table}[t]
    \centering
    \caption{Elementary statistics of the 10 real-world directed networks.}
    \label{tab:dataset_statistics_compact}
    \setlength{\tabcolsep}{10pt}
    \renewcommand{\arraystretch}{1.06}
    \begin{tabular}{lrrr}
        \toprule
        Network & $|V|$ & $|E|$ & $|E|/|V|$ \\
        \midrule
        Twitter & 81,306 & 1,768,149 & 21.75 \\
        Google+ & 201,949 & 1,496,936 & 7.41 \\
        PGP Web of Trust & 39,796 & 301,498 & 7.58 \\
        Facebook Wall Posts & 43,953 & 262,631 & 5.98 \\
        Epinions & 75,877 & 508,836 & 6.71 \\
        Digg Friends & 261,489 & 2,232,942 & 8.54 \\
        Weibo Followers & 88,532 & 8,174,684 & 92.34 \\
        Munmun Twitter & 465,017 & 834,797 & 1.80 \\
        YouTube Social Links & 1,134,885 & 4,938,950 & 4.35 \\
        Enron Email & 67,521 & 273,649 & 4.05 \\
        \bottomrule
    \end{tabular}
\end{table}

For the baseline implementations, Degree uses in-degree~\cite{freeman1978centrality}, PageRank uses a damping factor of $0.85$~\cite{page1999pagerank}, Katz uses an attenuation parameter of $0.005$~\cite{katz1953status}, Betweenness is approximated on the reversed directed graph with 200 sampled pivots~\cite{brandes2007centrality}, LC uses an in-degree-based directed adaptation~\cite{joyce2010leverage}, VoteRank uses an incremental directed adaptation on the reversed graph~\cite{zhang2016voterank}, LeaderRank is implemented using a bidirectionally connected ground node and PageRank with damping factor $0.85$~\cite{lu2011leaderrank}, CI uses radius 2 and a fixed in-degree-based ranking without iterative node removal or reranking~\cite{morone2015influence}.
\begin{table*}[t]
    \centering
    \footnotesize
    \setlength{\tabcolsep}{3pt}
    \renewcommand{\arraystretch}{1.15}
    \newcommand{\best}[1]{\textbf{#1}}

    \caption{Mean fraction of ever-infected nodes over 100 runs. Lower is better. Bold values indicate the best method for each network.}
    \label{tab:primary_final_infection}

    \begin{adjustbox}{max width=\textwidth}
    \begin{tabular}{@{}l*{10}{r}@{}}
        \toprule
        Method
        & Twitter & Google+
        & PGP & Facebook & Epinions
        & Digg & Weibo & Munmun & YouTube & Enron \\
        \midrule

        Degree
        & 0.07650 & 0.08061 & 0.07258 & 0.11417 & 0.01824
        & 0.01668 & 0.02013 & 0.01061 & 0.01677 & \best{0.01492} \\

        PageRank
        & 0.07484 & 0.04505 & 0.07100 & 0.12338 & 0.01844
        & 0.01625 & 0.02012 & 0.01103 & 0.01643 & 0.01528 \\

        Katz
        & 0.08164 & 0.12021 & 0.08915 & 0.11934 & 0.01871
        & 0.02192 & 0.02155 & 0.01061 & 0.01959 & 0.01535 \\

        Betweenness
        & 0.10104 & 0.11726 & 0.06537 & 0.14942 & 0.02274
        & 0.02019 & 0.04771 & 0.01154 & 0.01712 & 0.01560 \\

        LeaderRank
        & 0.07408 & 0.04209 & 0.07141 & 0.11078 & 0.01839
        & 0.01663 & 0.02006 & 0.01097 & 0.01632 & 0.01531 \\

        VoteRank
        & 0.07476 & 0.03926 & 0.06596 & 0.11006 & \best{0.01792}
        & 0.01558 & 0.02008 & 0.01063 & 0.01593 & 0.01506 \\

        CI
        & 0.07472 & 0.11119 & 0.08656 & 0.14966 & 0.01820
        & 0.01697 & 0.02022 & 0.01066 & 0.01733 & 0.01532 \\

        LC
        & 0.07405 & 0.05763 & 0.15813 & 0.11012 & 0.01835
        & 0.01662 & 0.04613 & 0.01074 & 0.01635 & 0.01640 \\

        \midrule

        HMI-Degree
        & 0.06824 & 0.03349 & 0.06341 & 0.10203 & 0.01812
        & \best{0.01557} & 0.01940 & \best{0.01047} & 0.01587 & 0.01502 \\

        HMI-PageRank
        & 0.06623 & 0.03291 & 0.06350 & 0.10345 & 0.01806
        & 0.01564 & 0.01938 & 0.01049 & 0.01591 & 0.01503 \\

        HMI-Katz
        & 0.06740 & 0.11993 & 0.06426 & 0.10522 & 0.01830
        & 0.02192 & \best{0.01932} & \best{0.01047} & 0.01959 & 0.01495 \\

        HMI-Betweenness
        & \best{0.06619} & \best{0.03286} & \best{0.06251}
        & \best{0.09942} & 0.01808
        & 0.01563 & 0.01936 & \best{0.01047} & \best{0.01586} & 0.01521 \\

        \bottomrule
    \end{tabular}
    \end{adjustbox}
\end{table*}

In the experiments, we randomly select $0.01|V|$ initially infected nodes, and each method selects $K=0.001|V|$ debunkers (both counts rounded down to integers). All infected nodes and selected debunkers are activated at $t=0$. The DA-SIR parameters are set to $\beta=0.15$, $\gamma=0.10$, $\alpha=0.20$, and $\tau=2$. For each network, the debunker sets selected by all twelve methods are determined before sampling the initially infected nodes. Results are averaged over 100 independent runs.

\begin{figure*}[p]
    \centering
    \includegraphics[width=\textwidth]{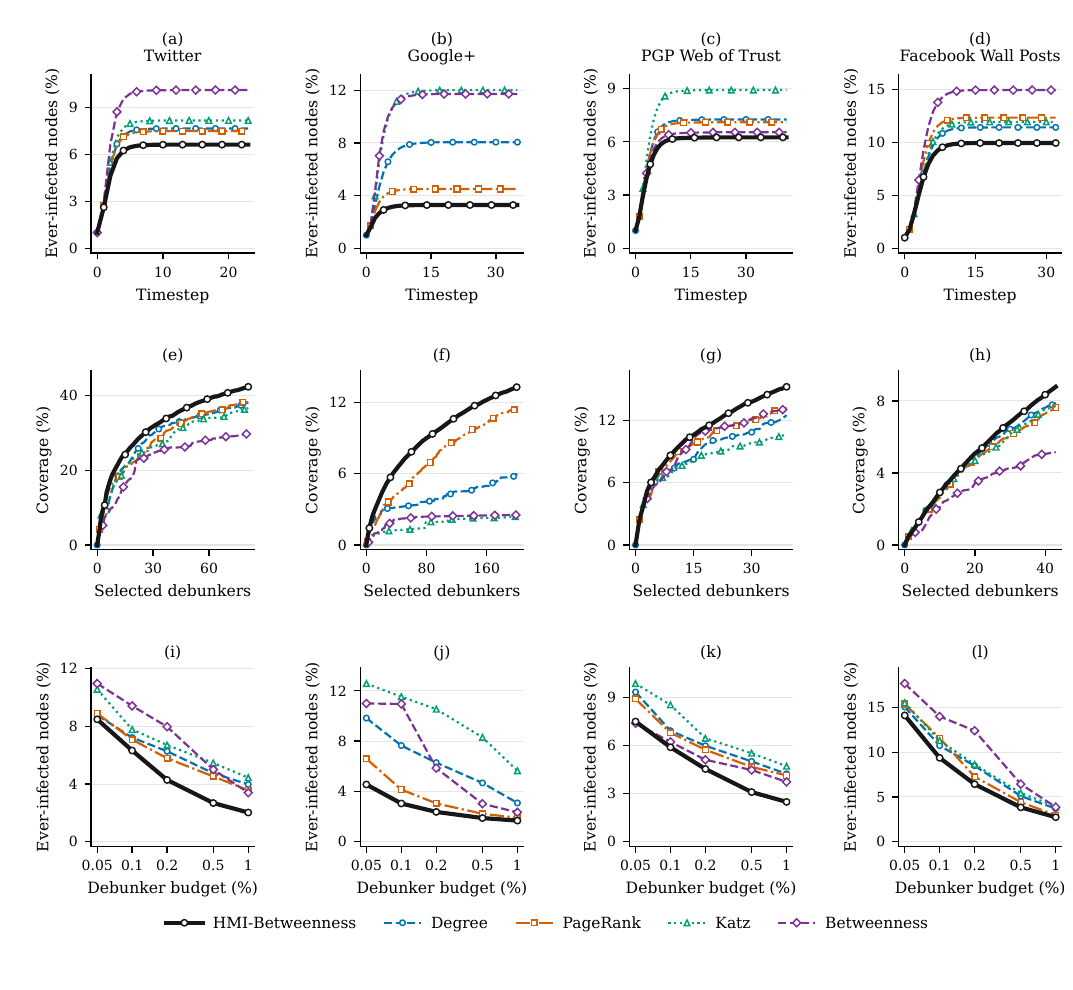}
    \caption{Rumor spreading, audience coverage, and budget dependence on Twitter, Google+, PGP, and Facebook Wall Posts (columns, left to right). (a)--(d) Percentage of ever-infected nodes over time. (e)--(h) Cumulative direct audience coverage. (i)--(l) Final percentage of ever-infected nodes versus debunker budget on a logarithmic scale. HMI-Betweenness (bold black) is compared with Degree, PageRank, Katz, and Betweenness. Infection results are averaged over 100 runs.}
    \label{fig:combined_results}
\end{figure*}

We first compare the methods in suppressing rumor spreading. The primary metric is the fraction of ever-infected nodes. Table~\ref{tab:primary_final_infection} reports the mean values for all twelve methods on ten networks. HMI achieves a lower mean than its corresponding base-centrality method in 37 of 40 matched comparisons. Among all twelve methods, an HMI variant achieves the lowest mean on eight of the ten networks. The first row of Fig.~\ref{fig:combined_results} shows the fraction of ever-infected nodes over time on four representative networks, where HMI-Betweenness achieves the lowest final fraction of ever-infected nodes among the displayed methods.

To better understand the differences in spreading outcomes, we next examine the direct audience coverage of the selected debunkers. Given the selected debunkers in order $v_1,\ldots,v_K$, the cumulative coverage after the first $k$ selections is defined as
\begin{equation}
\mathrm{Coverage}(k)
=\frac{1}{|V|}\left|\bigcup_{i=1}^{k}\mathcal{N}(v_i)\right|,
\qquad \mathrm{Coverage}(0)=0.
\label{eq:results_coverage}
\end{equation}
This measure quantifies the direct reach of the selected debunkers. The second row of Fig.~\ref{fig:combined_results} shows cumulative coverage on four representative networks. HMI-Betweenness reaches more distinct nodes with the same number of selected debunkers. Across all ten networks and four corresponding base-centrality methods, HMI achieves larger final coverage, $\mathrm{Coverage}(K)$, in 38 of 40 matched comparisons. 

We further evaluate HMI under different intervention budgets, intervention times, and spreading parameters. The third row of Fig.~\ref{fig:combined_results} shows the effect of increasing the debunker budget from $0.05\%$ to $1\%$. The mean fraction of ever-infected nodes decreases as the budget increases for all methods. HMI achieves a lower mean than its corresponding centrality method in 192 of 200 matched comparisons. With intervention times $t_d\in\{0,1,2\}$, HMI achieves a lower mean fraction of ever-infected nodes in 37, 34, and 32 of 40 matched comparisons, respectively. Across five rumor-transmission settings ($\beta$ from $0.05$ to $0.25$), HMI achieves lower means in 185 of 200 comparisons. Across three corrective-information settings, with $(\alpha,\gamma)$ ranging from $(0.10,0.05)$ to $(0.30,0.15)$, HMI achieves lower means in 109 of 120 comparisons.

\section{Discussion}
\label{sec:discussion}

Our results highlight an important distinction between the structural importance of individual nodes and their collective contributions. In contrast to conventional centrality-based methods that rely on fixed node rankings regardless of previous selections, HMI evaluates each candidate relative to the selected debunkers, dynamically updating its marginal contribution throughout the selection process. By accounting for already reached audiences while retaining centrality information, HMI improves both direct audience coverage and rumor suppression. This effect is particularly evident for Betweenness centrality, which performs relatively poorly as a standalone heuristic on several networks but achieves substantially better results when combined with marginal coverage. Notably, HMI-Betweenness attains the lowest mean final fraction of ever-infected nodes among all twelve methods on Twitter, Google+, PGP, Facebook Wall Posts, and YouTube. These findings show that incorporating marginal coverage can substantially improve the effectiveness of a centrality measure even when its original ranking performs poorly.

Rather than simply maximizing the number of distinct nodes covered, HMI accounts for the potential benefits of repeated exposure to corrective information by assigning diminishing yet positive weights to multiple coverages. This formulation favors previously uncovered audiences without completely disregarding additional exposures, while the dispersion-based weighting mechanism adjusts the balance between centrality and marginal coverage. Together, these mechanisms enable HMI to balance individual structural importance and collective audience reach, without requiring spreading parameters.

The intervention-budget experiments further underscore the importance of accounting for previously selected debunkers as the selected set expands. While conventional centrality methods retain their original rankings, HMI continuously updates the marginal contributions of remaining candidates, achieving a lower fraction of ever-infected nodes in all 40 matched comparisons at the largest budget. The improvements observed across different intervention times and spreading parameters further demonstrate the effectiveness of coverage-aware selection under varying spreading conditions.

\bibliography{references}
\end{document}


\title{\texorpdfstring{Supplemental Material for\\ Locating Debunkers for Rumor Mitigation in Directed Networks}{Supplemental Material for Locating Debunkers for Rumor Mitigation in Directed Networks}}
\maketitle

\renewcommand{\thesection}{S\arabic{section}}
\renewcommand{\thesubsection}{\thesection.\arabic{subsection}}
\makeatletter
\renewcommand{\p@subsection}{}
\makeatother
\renewcommand{\theequation}{S\arabic{equation}}
\renewcommand{\thefigure}{S\arabic{figure}}
\renewcommand{\thetable}{S\arabic{table}}

\section{Time and space complexity of HMI}
\label{app:hmi_complexity}

We analyze the time and space complexity of HMI on a network with \(N\) nodes and debunker budget \(1\leq K\leq N\). Let \(T_c\) and \(S_c\) denote the time and space costs of the chosen base centrality. Let
\[
M=\sum_{v\in V}|\mathcal N(v)|=Ns,
\qquad
a(u)=|\{v:u\in\mathcal N(v)\}|,
\]
where \(s\) is the average neighborhood size and \(a(u)\) is the length of the inverted list for user \(u\). Neighborhoods include their candidate, so \(M\geq N\). For a directed graph without parallel edges, \(M=N+|E|-L\), where \(L\) is the number of self-loops (a loop does not add a second copy of its node). We assume constant-time arithmetic and expected constant-time dictionary and set operations.

Constructing all coverage neighborhoods and the inverted index takes \(O(M)\) time and space. Graph-topology validation takes \(O(N+|E|)\), also \(O(M)\) for these coverage neighborhoods. Initial gains follow from neighborhood sizes. The implementation additionally sorts all candidates by decreasing base centrality to establish their stable order, costing \(O(N\log N)\) time and \(O(N)\) space even when the base scores are already cached.

The Jeffreys cache starts at \(w_0=1\) and generates each later weight once using Eq.~\eqref{main-eq:jeffreys_recurrence} of the main text. Coverage counts cannot exceed \(K\), so the cache costs \(O(K)\) total time and space. Each new entry requires constant-time arithmetic.

At each step, dynamic min--max normalization, the two normalized standard deviations, adaptive mixing, and the maximum-score scan cost \(O(N)\). Selecting \(v\) increments coverage counts for users in \(\mathcal N(v)\). Each changed weight is propagated through its inverted list, giving the update cost
\[
O\!\left(\sum_{u\in\mathcal N(v)}a(u)\right).
\]
This counts visits even to list entries whose candidates have already been selected; their gains are skipped. Updating the affected-gain array adds no larger term. New weight generation is amortized over the run.

For the final selected set \(D_K\), the bound is
\[
\boxed{
\begin{aligned}
T=T_c+O\!\Bigl(&N\log N+M+KN\\
&+\sum_{v\in D_K}\sum_{u\in\mathcal N(v)}a(u)\Bigr).
\end{aligned}
}
\]
If selected neighborhoods have size \(O(s)\) and their covered users have inverted lists of length \(O(s)\), this simplifies to
\[
T=T_c+O\big(N\log N+Ns+K(N+s^2)\big).
\]
The simplified bound requires both stated neighborhood conditions.

With uniformly bounded neighborhoods and inverted lists, selection excluding the base computation takes \(O(N\log N+KN)\). For dense neighborhoods, each update can cost \(O(N^2)\), giving the worst-case selection bound \(O(KN^2)\).

Neighborhoods and inverted lists occupy \(O(M)\), candidate arrays and coverage counts \(O(N)\), and the weight cache and selected-node records \(O(K)\). Thus the default selector has
\[
\boxed{S=S_c+O(M+K)=S_c+O(Ns+K).}
\]
If the optional full candidate-gain trace is stored, it adds \(\Theta(KN)\) time and space. This trace is excluded from the default bound.

\section{Additional Experimental Results}
\label{app:additional_experiments}

The main results retain every fixed HMI variant on every network. This
section addresses the remaining questions: whether the advantage extends
to other baselines and epidemic outcomes, where it fails under changed
conditions, and what the selection process and component ablation reveal.
Tables summarize the full evaluation grids; illustrative plots expose
contrasting behaviors. Complete numerical trajectories and per-step
diagnostics remain in the accompanying evidence archive.

\subsection{Graph structure and complete primary comparison}
\label{app:detailed_dataset_statistics}
\label{app:all_methods}

Table~\ref{tab:compact_graph_statistics} complements the graph sizes in the
main text. The ten cached graphs are weakly connected but differ
substantially in strong connectivity and clustering. Twitter and Google+
retain 14 and 1,713 self-loops; those arcs remain in diffusion and coverage
calculations. Only the simple undirected projection used for triangle and
clustering statistics removes them. These heterogeneous graph properties
provide context for the network-dependent results, rather than establishing
a causal explanation for a particular method's advantage.

\begin{table*}[t]
\centering
\small
\setlength{\tabcolsep}{4pt}
\caption{Additional graph statistics. SCC is the percentage of nodes in the largest strongly connected component; node and arc totals are in Table~\ref{main-tab:dataset_statistics_compact} of the main text. Clustering, triangles and transitivity use the simple undirected projection. All processed graphs are weakly connected.}
\label{tab:compact_graph_statistics}
\begin{tabular}{lrrrrr}
\toprule
Network & SCC (\%) & Loops & Clustering & Triangles & Transitivity \\
\midrule
Twitter & 84.1 & 14 & 0.5653 & 13,082,506 & 0.1706 \\
Google+ & 43.0 & 1713 & 0.1479 & 12,196,058 & 0.2388 \\
PGP & 100.0 & 0 & 0.4611 & 1,146,500 & 0.2538 \\
Facebook & 70.1 & 0 & 0.1149 & 122,842 & 0.0851 \\
Epinions & 42.5 & 0 & 0.1378 & 1,624,481 & 0.0657 \\
Digg & 50.5 & 0 & 0.0969 & 14,236,164 & 0.0614 \\
Weibo & 67.4 & 0 & 0.1349 & 82,688,534 & 0.0164 \\
Munmun Twitter & 0.4 & 0 & 0.0152 & 38,389 & 0.0006 \\
YouTube & 44.9 & 0 & 0.0808 & 3,049,268 & 0.0062 \\
Enron & 12.2 & 0 & 0.1398 & 1,067,980 & 0.0736 \\
\bottomrule
\end{tabular}
\end{table*}

\begin{table*}[t]
    \centering
    \footnotesize
    \setlength{\tabcolsep}{3pt}
    \caption{Population standard deviations of the final ever-infected percentages over 100 paired seeds for the eight methods in Table~\ref{main-tab:primary_final_infection} of the main text. Values are in percentage points; the corresponding means are reported in the main text.}
    \label{tab:primary_standard_deviations}
    \begin{tabular}{l*{8}{r}}
        \toprule
        & \multicolumn{4}{c}{Base centralities} & \multicolumn{4}{c}{HMI variants} \\
        \cmidrule(lr){2-5}\cmidrule(lr){6-9}
        Network & Degree & PageRank & Katz & Betweenness & Degree & PageRank & Katz & Betweenness \\
        \midrule
        Twitter & $0.408$ & $0.390$ & $0.438$ & $0.527$ & $0.368$ & $0.342$ & $0.377$ & $0.382$ \\
        Google{+} & $0.307$ & $0.172$ & $0.353$ & $0.356$ & $0.129$ & $0.125$ & $0.312$ & $0.113$ \\
        PGP Web of Trust & $0.595$ & $0.574$ & $0.797$ & $0.463$ & $0.470$ & $0.479$ & $0.453$ & $0.459$ \\
        Facebook Wall Posts & $0.805$ & $0.868$ & $0.841$ & $1.025$ & $0.647$ & $0.768$ & $0.682$ & $0.661$ \\
        Epinions & $0.092$ & $0.092$ & $0.102$ & $0.173$ & $0.100$ & $0.092$ & $0.102$ & $0.086$ \\
        Digg Friends & $0.045$ & $0.039$ & $0.139$ & $0.057$ & $0.031$ & $0.033$ & $0.139$ & $0.029$ \\
        Weibo Followers & $0.071$ & $0.074$ & $0.101$ & $0.447$ & $0.071$ & $0.071$ & $0.071$ & $0.066$ \\
        Munmun Twitter & $0.003$ & $0.003$ & $0.003$ & $0.006$ & $0.003$ & $0.003$ & $0.003$ & $0.003$ \\
        YouTube Social Links & $0.021$ & $0.018$ & $0.038$ & $0.020$ & $0.017$ & $0.016$ & $0.038$ & $0.017$ \\
        Enron Email & $0.174$ & $0.170$ & $0.184$ & $0.168$ & $0.127$ & $0.141$ & $0.129$ & $0.124$ \\
        \bottomrule
    \end{tabular}
\end{table*}

\begin{table*}[t]
\centering
\small
\setlength{\tabcolsep}{4pt}
\caption{Additional primary baselines: final ever-infected percentage, mean $\pm$ population SD over 100 paired seeds. Together with Table~\ref{main-tab:primary_final_infection} of the main text, these entries retain the means for all twelve methods on all ten networks.}
\label{tab:appendix_all_baselines}
\begin{tabular}{lrrrr}
\toprule
Network & LeaderRank & VoteRank & CI & LC \\
\midrule
Twitter & $7.408\pm0.397$ & $7.476\pm0.398$ & $7.472\pm0.379$ & $7.405\pm0.432$ \\
Google+ & $4.209\pm0.147$ & $3.926\pm0.168$ & $11.119\pm0.257$ & $5.763\pm0.207$ \\
PGP & $7.141\pm0.508$ & $6.596\pm0.498$ & $8.656\pm0.761$ & $15.813\pm1.213$ \\
Facebook & $11.078\pm0.703$ & $11.006\pm0.722$ & $14.966\pm0.982$ & $11.012\pm0.797$ \\
Epinions & $1.839\pm0.107$ & $1.792\pm0.094$ & $1.820\pm0.094$ & $1.835\pm0.106$ \\
Digg & $1.663\pm0.040$ & $1.558\pm0.038$ & $1.697\pm0.043$ & $1.662\pm0.043$ \\
Weibo & $2.006\pm0.074$ & $2.008\pm0.073$ & $2.022\pm0.074$ & $4.613\pm0.571$ \\
Munmun Twitter & $1.097\pm0.003$ & $1.063\pm0.003$ & $1.066\pm0.003$ & $1.074\pm0.003$ \\
YouTube & $1.632\pm0.017$ & $1.593\pm0.016$ & $1.733\pm0.023$ & $1.635\pm0.019$ \\
Enron & $1.531\pm0.184$ & $1.506\pm0.125$ & $1.532\pm0.189$ & $1.640\pm0.215$ \\
\bottomrule
\end{tabular}
\end{table*}

Table~\ref{tab:appendix_all_baselines} adds LeaderRank, VoteRank, CI and LC
to the four raw bases and four HMI variants already reported in the main
text. Table~\ref{tab:primary_standard_deviations} provides the population
standard deviations for the eight methods in Table~\ref{main-tab:primary_final_infection} of the main text.
Thus means and standard deviations remain available for all twelve methods.
Table~\ref{tab:appendix_best_family_comparison} gives the paired differences
for the descriptive family comparison. It preserves the small Epinions and
Enron deficits and the unresolved Digg difference.
Choosing both methods by their observed means makes these intervals
exploratory; they neither correct for selection nor validate a deployment
rule for choosing HMI's base.

\begin{table*}[t]
    \centering
    \footnotesize
    \setlength{\tabcolsep}{3.5pt}
    \caption{Descriptive best-of-family comparison. The paired difference is the best-baseline infection percentage minus the best-HMI infection percentage; positive values favor HMI. The interval is the ordinary seed-wise 95\% $t$ interval after selecting both methods by their campaign means and is therefore descriptive rather than selection-adjusted.}
    \label{tab:appendix_best_family_comparison}
    \begin{tabular}{lllrr}
        \toprule
        Network & Best HMI & Best baseline & Difference (pp) & Relative change \\
        \midrule
        Twitter & HMI-Betweenness & LC & $0.786 [0.724, 0.847]$ & $+10.61\%$ \\
        Google{+} & HMI-Betweenness & VoteRank & $0.640 [0.606, 0.674]$ & $+16.30\%$ \\
        PGP Web of Trust & HMI-Betweenness & Betweenness & $0.286 [0.186, 0.387]$ & $+4.38\%$ \\
        Facebook Wall Posts & HMI-Betweenness & VoteRank & $1.064 [0.904, 1.223]$ & $+9.67\%$ \\
        Epinions & HMI-PageRank & VoteRank & $-0.014 [-0.029, 0.002]$ & $-0.77\%$ \\
        Digg Friends & HMI-Degree & VoteRank & $0.001 [-0.006, 0.008]$ & $+0.07\%$ \\
        Weibo Followers & HMI-Katz & LeaderRank & $0.074 [0.066, 0.083]$ & $+3.70\%$ \\
        Munmun Twitter & HMI-Katz & Katz & $0.014 [0.013, 0.015]$ & $+1.34\%$ \\
        YouTube Social Links & HMI-Betweenness & VoteRank & $0.006 [0.003, 0.009]$ & $+0.38\%$ \\
        Enron Email & HMI-Katz & Degree & $-0.003 [-0.030, 0.023]$ & $-0.23\%$ \\
        \bottomrule
    \end{tabular}
\end{table*}

\begin{figure*}[t]
\centering
\includegraphics[width=.88\textwidth]{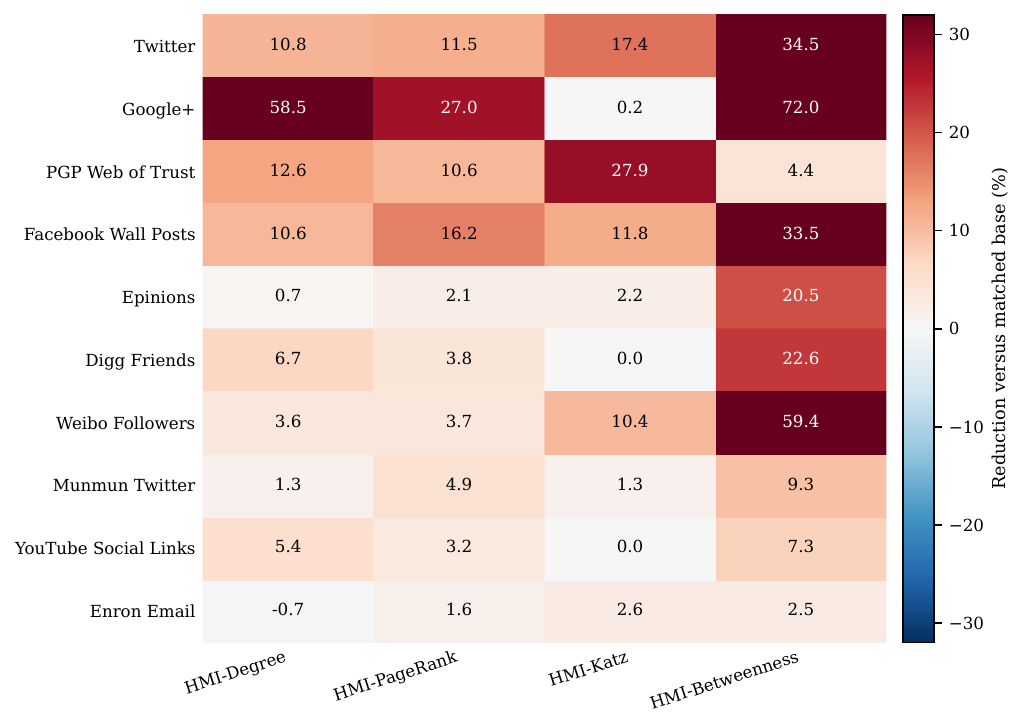}
\caption{Relative reduction in mean final infections from each raw base to
its fixed HMI variant on all ten networks. Positive values favor HMI.
This complete matrix separates the benefit of hybridization from the
descriptive choice of the best HMI base; it counts differences in means,
not statistically significant wins.}
\label{fig:compact_matched_effects}
\end{figure*}

\paragraph{Matched-base effects.}
\label{app:hmi_versatility}
Figure~\ref{fig:compact_matched_effects} displays all 40 fixed HMI--base
comparisons. Large relative reductions occur where a raw centrality is
poor at collective placement; they should not be confused with the smaller
reductions against the strongest of all eight baselines. HMI-Katz remains
close to raw Katz on Google+, Digg and YouTube, while the benefit from
Betweenness changes markedly across networks. The only adverse primary
matched mean is HMI-Degree on Enron, with an ordinary paired interval crossing
zero. Of the 37 lower means, 34 pass the two-sided paired test after Holm
correction \cite{holm1979simple} across all 40 comparisons: eight for HMI-Degree, nine for
HMI-PageRank, seven for HMI-Katz and ten for HMI-Betweenness.
The three lower means that do not pass are HMI-Degree on Epinions
($p_{\mathrm H}=1.000$), HMI-Katz on Google+ ($1.000$), and HMI-PageRank
on Enron ($0.259$).
The higher HMI-Degree mean on Enron also does not pass ($p_{\mathrm H}=1.000$).
The two exact observed ties are retained as ties, without an equivalence
claim. The statistical archive reports all 40 effects, pointwise intervals,
unadjusted test values and Holm-adjusted values.

\subsection{Secondary outcomes and informative trajectories}
\label{app:restored_secondary}
\label{app:complete_trajectories}
\label{app:restored_trajectories}

Lower final infection is usually accompanied by lower concurrent burden.
Across the 40 fixed comparisons, HMI has a lower peak in 36, a tie in two,
and a higher peak in two; the AUC counts are 37, two and one.
Table~\ref{tab:compact_secondary} separates these outcomes by base across all
ten networks. With Holm correction over the 80 peak/AUC comparisons
jointly, 32 peak reductions and 34 AUC reductions remain significant.
All adverse secondary means occur on Enron. HMI-Degree's peak increase is small
but significant: $+0.0130$ percentage points, pointwise 95\% interval
$[0.0083,0.0177]$, $p_{\mathrm H}=5.96\times10^{-6}$.
HMI-Katz's peak increase of $0.0064$ points does not pass the corrected test
($p_{\mathrm H}=0.170$), nor does HMI-Degree's AUC increase of
$8.16\times10^{-5}$ timesteps ($p_{\mathrm H}=1.000$).
These results separate a detectable adverse peak effect from its small
absolute magnitude; lower final infections need not imply improvement in
every secondary endpoint.
Peak timing has no universal favorable direction, so earlier peaks are not
counted as an additional containment benefit.

\begin{table*}[t]
\centering
\small
\setlength{\tabcolsep}{4pt}
\caption{Primary HMI--matched-base comparisons across all ten networks. Each cell is the number of lower/equal/higher means. Lower final infections, peak and AUC indicate less rumor burden; a larger correction counter is not a count of distinct protected users. These are descriptive counts, not significance tests.}
\label{tab:compact_secondary}
\begin{tabular}{lcccc}
\toprule
HMI base & Final infections & Peak infections & Infection AUC & Correction counter \\
\midrule
Degree & 9/0/1 & 9/0/1 & 9/0/1 & 0/0/10 \\
PageRank & 10/0/0 & 10/0/0 & 10/0/0 & 0/0/10 \\
Katz & 8/2/0 & 7/2/1 & 8/2/0 & 1/2/7 \\
Betweenness & 10/0/0 & 10/0/0 & 10/0/0 & 2/0/8 \\
\bottomrule
\end{tabular}
\end{table*}

\begin{figure*}[t]
\centering
\includegraphics[width=\textwidth]{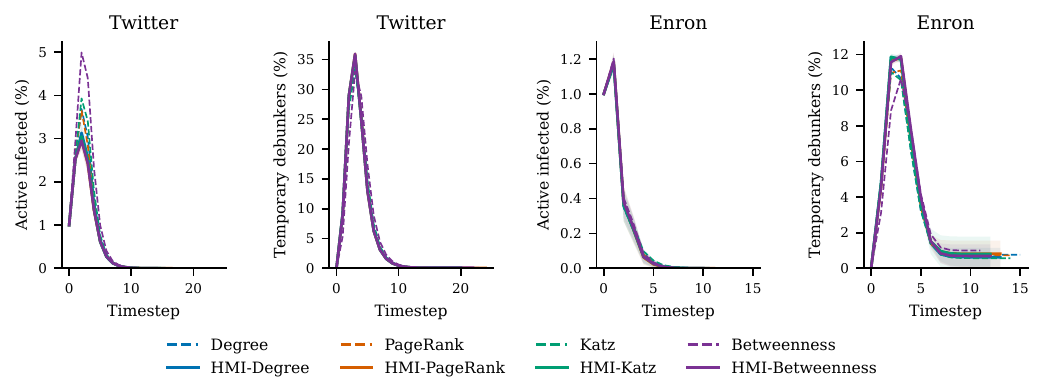}
\caption{Contrasting active-infection and temporary-debunker trajectories
on Twitter and Enron. Solid/dashed lines of the same color form a fixed
HMI--base pair. Lines are means over 100 common seeds; HMI shading shows
one population SD. Final recorded values are carried to a common display
horizon. Axes vary by panel. The cases illustrate substantial benefits
on Twitter and near parity on Enron, including the adverse HMI-Degree mean.
Full comparisons remain in the numerical tables.}
\label{fig:compact_trajectories}
\end{figure*}

Figure~\ref{fig:compact_trajectories} contrasts Twitter, where the primary
matched effects are substantial, with Enron, where they are small and
HMI-Degree has the adverse mean. The plots show active rumor and temporary
corrective-information propagation for all four fixed HMI variants and their matched
bases. They complement the four-network cumulative trajectories in the
main text; their purpose is to illustrate distinct temporal behavior, not
to estimate a network-average effect from two selected examples. All
twelve-method trajectories on all ten graphs remain in the evidence
archive.

\paragraph{Epidemic-state interpretation.}
\label{app:state_diagnostics}
The five state occupancies sum to the graph population at every recorded
step in all 12,000 primary runs, as checked in the saved trajectory audit. The recovered and susceptible curves therefore help
check the simulator but add little independent evidence about which
selector is best once infection burden and corrective-information propagation are shown.
Selected debunkers propagate corrective information only during their specified active spreading period.

The recorded correction counter is the sum of two population fractions:
users ever recovered directly by debunking and users ever becoming temporary
debunkers. These sets can
overlap, and their members may have been infected earlier. A larger counter
therefore cannot be read as more distinct users protected from infection.
Its complete values remain in the secondary-outcome CSV; the interpretation applies equally to every saved state/counter trajectory.

\subsection{Sensitivity to propagation and intervention settings}
\label{app:sensitivity}

The one-factor sweeps evaluate four HMIs and their four matched raw bases
on all ten graphs. Delay uses $t_d\in\{0,1,2\}$, transmission uses
$\beta\in\{0.05,0.10,0.15,0.20,0.25\}$, and correction uses weak,
canonical and strong $(\alpha,\gamma)$ pairs of $(0.10,0.05)$,
$(0.20,0.10)$ and $(0.30,0.15)$. These sweeps hold the selected nodes and
initial sources fixed and reuse their canonical primary outcomes.
Table~\ref{tab:compact_sweep_counts} summarizes every base and setting;
Table~\ref{tab:compact_sweep_exceptions} identifies every adverse
network--base group and its exact settings. Repeated canonical points and
correlated settings are not independent replications or significance tests.

\begin{table}[t]
\centering
\small
\setlength{\tabcolsep}{4pt}
\caption{All-network sweep summary: lower/equal/higher mean final infections relative to each fixed raw base. Each base has 50 budget, 30 delay, 50 transmission and 30 correction comparisons. Canonical points recur across three sweeps; the counts are dependent descriptive comparisons.}
\label{tab:compact_sweep_counts}
\begin{tabular}{lcccc}
\toprule
HMI base & Budget & Delay & Transmission & Correction \\
\midrule
Degree & 47/0/3 & 27/0/3 & 47/0/3 & 28/0/2 \\
PageRank & 49/0/1 & 28/0/2 & 50/0/0 & 29/0/1 \\
Katz & 48/2/0 & 21/6/3 & 38/10/2 & 23/6/1 \\
Betweenness & 48/0/2 & 27/0/3 & 50/0/0 & 29/0/1 \\
\bottomrule
\end{tabular}
\end{table}

\begin{table*}[t]
\centering
\small
\setlength{\tabcolsep}{4pt}
\caption{Every network--base group with an adverse final-infection mean in any sweep. The settings enumerate all higher-mean cells, including reused canonical cells. Maximum excess is HMI minus its matched base in percentage points; small excesses need not be statistically resolved.}
\label{tab:compact_sweep_exceptions}
\begin{tabular}{llp{0.40\textwidth}r}
\toprule
Network & HMI base & Settings with a higher mean & Max. excess (pp) \\
\midrule
Google+ & Katz & $\beta$: 0.05, 0.2 & 0.00445 \\
PGP & Betweenness & Budget (\%): 0.05 & 0.11054 \\
Munmun Twitter & Katz & Delay: 2 & 0.00026 \\
Munmun Twitter & Betweenness & Delay: 2 & 0.07431 \\
Enron & Degree & Budget (\%): 0.1, 0.2, 0.5; Delay: 0, 1, 2; $\beta$: 0.15, 0.1, 0.2; Correction: canonical, weak & 0.96180 \\
Enron & PageRank & Budget (\%): 0.5; Delay: 1, 2; Correction: weak & 1.14587 \\
Enron & Katz & Delay: 1, 2; Correction: weak & 0.92989 \\
Enron & Betweenness & Budget (\%): 0.1; Delay: 1, 2; Correction: weak & 1.40998 \\
\bottomrule
\end{tabular}
\end{table*}

Mean final infections increase with delay and transmission, and decrease
with stronger correction, for all 80 network--method curves in each sweep.
This uniform direction concerns the behavioral intervention parameters;
HMI's advantage over its own base is less uniform. The delay counts favor
HMI in 37 of 40 comparisons at $t_d=0$, 34 at $t_d=1$, and 32 at $t_d=2$,
with two ties at each delay. Transmission has 185 lower HMI means, ten
ties and five higher means over 200 comparisons. Correction has 109 lower,
six equal and five higher means over 120 comparisons.

\begin{figure*}[t]
\centering
\includegraphics[width=\textwidth]{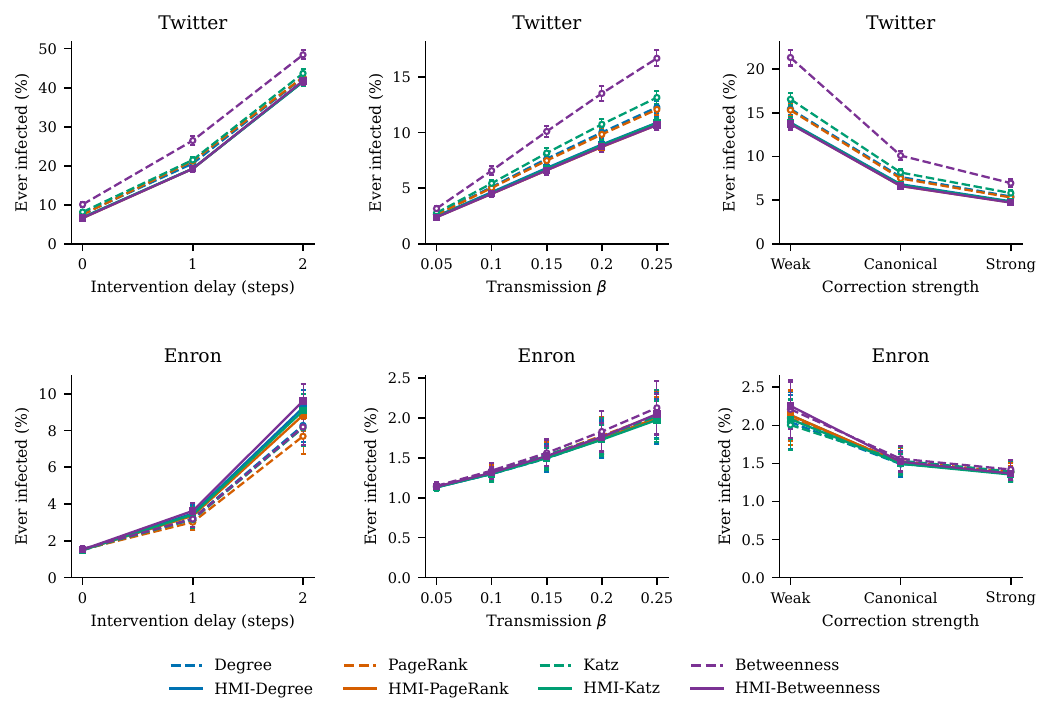}
\caption{Sensitivity contrasts on Twitter and Enron, using all four fixed
HMI variants and matched raw bases. Rows show Twitter and Enron; columns vary delay, transmission and correction; all other settings are fixed. Points are 100-seed means and
error bars are population SDs, not confidence intervals. The examples were
chosen to show both a substantial matched benefit and systematic delayed
intervention failures; Tables~\ref{tab:compact_sweep_counts}
and~\ref{tab:compact_sweep_exceptions} cover all ten networks.}
\label{fig:compact_sensitivity}
\end{figure*}

The contrast in Figure~\ref{fig:compact_sensitivity} is consequential.
On Twitter, HMI-Betweenness rises from $6.619\%$ final infection with
immediate intervention to $19.255\%$ after one step and $41.747\%$ after
two: good placement does not offset late intervention. On Enron, all four
HMI variants have higher means than their raw bases at delays one and two;
the largest excess is 1.410 percentage points for HMI-Betweenness at two
steps. This failure would be obscured by showing only the strongest
primary-network gains. Google+ Katz and delayed Munmun
Twitter also contribute adverse cells, usually much smaller in magnitude.

\subsection{Secondary outcomes across debunker budgets}
\label{app:budget_secondary}

The five budgets are $0.05\%$, $0.1\%$, $0.2\%$, $0.5\%$ and $1\%$.
All use the sweep-specific source pool described in
Section~\ref{main-sec:budget_results} of the main text; the primary $0.1\%$ result is not
inserted into this curve. Final infection decreases monotonically with
budget for all 80 network--method curves, and all 40 matched means favor
HMI at the largest budget. Across all budgets, the final-infection counts
are 192 lower, two equal and six higher. The adverse settings in
Table~\ref{tab:compact_sweep_exceptions} include PGP HMI-Betweenness at
$0.05\%$ and several Enron cases; a favorable overall count does not
remove these exceptions.

\begin{table*}[t]
\centering
\small
\setlength{\tabcolsep}{4pt}
\caption{Secondary budget outcomes over all ten networks and five budgets: lower/equal/higher means versus the matched base, 50 comparisons per cell. The same source pool is used across all five budgets; these counts are descriptive and correlated.}
\label{tab:compact_budget_secondary}
\begin{tabular}{lccc}
\toprule
HMI base & Peak infections & Infection AUC & Correction counter \\
\midrule
Degree & 43/3/4 & 48/0/2 & 2/0/48 \\
PageRank & 47/3/0 & 50/0/0 & 0/0/50 \\
Katz & 42/5/3 & 48/2/0 & 10/2/38 \\
Betweenness & 48/2/0 & 48/0/2 & 9/0/41 \\
\bottomrule
\end{tabular}
\end{table*}

Table~\ref{tab:compact_budget_secondary} summarizes peak infections, AUC
and the correction counter across every tested budget and graph. These
quantities distinguish concurrent rumor burden from its cumulative
duration and corrective activity. They use fixed base identities rather
than choosing the best HMI separately at each budget. The complete
per-setting means and population SDs remain in the sweep CSV, allowing
inspection of individual curves including seed-to-seed variability and
coincident curves. These SDs are not standard errors of paired differences;
the sweep summaries alone do not establish the significance of matched
effects.

\begin{table*}[t]
\centering
\small
\setlength{\tabcolsep}{3pt}
\caption{Mean shared directly reachable nodes per debunker pair for the fixed
primary selections (Base versus HMI). HMI reduces overlap in 28 of 30
comparisons for Degree, PageRank, and Katz (two ties); Betweenness
instead shows seven increases and three decreases. Those seven increases
accompany larger mean individual audiences and broader union coverage,
illustrating that greater absolute overlap can coexist with greater
collective reach.}
\label{tab:audience_overlap}
\begin{tabular}{lrrrrrrrr}
\toprule
 & \multicolumn{2}{c}{Degree} & \multicolumn{2}{c}{PageRank}
 & \multicolumn{2}{c}{Katz} & \multicolumn{2}{c}{Betweenness} \\
\cmidrule(lr){2-3}\cmidrule(lr){4-5}\cmidrule(lr){6-7}\cmidrule(lr){8-9}
Network & Base & HMI & Base & HMI & Base & HMI & Base & HMI \\
\midrule
Twitter & 82.16 & 54.53 & 60.65 & 48.89 & 90.70 & 54.69 & 23.45 & 48.34 \\
Google+ & 66.27 & 3.42 & 2.91 & 1.84 & 115.05 & 114.24 & 0.04 & 1.83 \\
PGP & 44.55 & 15.70 & 25.75 & 14.44 & 59.95 & 16.63 & 15.66 & 11.90 \\
Facebook & 2.52 & 1.46 & 1.21 & 1.19 & 2.75 & 1.53 & 0.52 & 1.05 \\
Epinions & 94.00 & 85.83 & 88.23 & 77.53 & 99.05 & 94.39 & 16.44 & 62.64 \\
Digg & 238.39 & 55.52 & 118.63 & 44.30 & 291.67 & 291.67 & 8.35 & 42.81 \\
Weibo & 3679.31 & 3263.66 & 3482.39 & 3222.62 & 3291.88 & 3239.66 & 238.15 & 3169.63 \\
Munmun Twitter & 4.94 & 2.68 & 4.43 & 1.69 & 4.93 & 2.70 & 0.06 & 1.65 \\
YouTube & 8.88 & 5.29 & 5.43 & 5.11 & 11.19 & 11.19 & 5.60 & 5.05 \\
Enron & 51.18 & 18.62 & 31.68 & 17.60 & 56.06 & 20.18 & 20.90 & 13.48 \\
\bottomrule
\end{tabular}
\end{table*}

\subsection{HMI selection dynamics and structural coverage}
\label{app:hmi_selection}
\label{app:restored_coverage}

The selection diagnostics test whether the hybrid changes placement and
how its mixing weight behaves. The 40 selector-only replays reproduce all
saved primary node orders exactly, covering 9,816 selection steps.
Independent gain recomputation, centrality-cache checks and hybrid-score
identities validate the traces. No diffusion was rerun to obtain them.

\begin{table*}[t]
\centering
\small
\setlength{\tabcolsep}{4pt}
\caption{Matched structural comparison for all four HMI variants: direct audience coverage gain over the raw base in percentage points, followed by selected-set Jaccard overlap in brackets. Zero gain or unit overlap exposes cases where hybridization changes little. The diagnostic excludes automatic self-coverage.}
\label{tab:compact_coverage}
\begin{tabular}{lrrrr}
\toprule
Network & Degree & PageRank & Katz & Betweenness \\
\midrule
Twitter & $+3.828\;[0.588]$ & $+3.969\;[0.473]$ & $+5.321\;[0.486]$ & $+12.214\;[0.191]$ \\
Google+ & $+7.082\;[0.149]$ & $+1.867\;[0.483]$ & $+0.075\;[0.990]$ & $+10.743\;[0.036]$ \\
PGP & $+2.493\;[0.418]$ & $+1.892\;[0.529]$ & $+4.375\;[0.300]$ & $+2.086\;[0.368]$ \\
Facebook & $+0.746\;[0.623]$ & $+0.944\;[0.623]$ & $+0.867\;[0.623]$ & $+3.624\;[0.178]$ \\
Epinions & $+0.805\;[0.786]$ & $+1.451\;[0.667]$ & $+1.178\;[0.744]$ & $+6.669\;[0.154]$ \\
Digg & $+3.454\;[0.286]$ & $+1.696\;[0.438]$ & $+0.000\;[1.000]$ & $+6.246\;[0.123]$ \\
Weibo & $+0.679\;[0.692]$ & $+0.637\;[0.676]$ & $+1.764\;[0.443]$ & $+10.278\;[0.080]$ \\
Munmun Twitter & $+0.066\;[0.378]$ & $+0.091\;[0.238]$ & $+0.066\;[0.382]$ & $+0.226\;[0.004]$ \\
YouTube & $+1.664\;[0.496]$ & $+0.537\;[0.726]$ & $+0.000\;[1.000]$ & $+2.644\;[0.356]$ \\
Enron & $+7.953\;[0.457]$ & $+6.123\;[0.506]$ & $+8.565\;[0.426]$ & $+12.358\;[0.196]$ \\
\bottomrule
\end{tabular}
\end{table*}

Table~\ref{tab:compact_coverage} combines two complementary diagnostics:
additional direct audience coverage and selected-set identity. Coverage
uses the union of directly reachable nodes and excludes automatic
self-contribution, whereas HMI's weighted neighborhood includes the
candidate itself. The main text defines its incremental form
$\Delta\mathrm{Coverage}_k$ and displays both cumulative and unsmoothed
incremental sequences for four networks in
Figure~\ref{main-fig:restored_main_coverage} of the main text. HMI has greater endpoint
coverage in 38 of 40 matched comparisons and equal coverage in two.
Yet HMI-Degree on Enron still has a higher primary infection mean. Broader
direct audience reach is therefore informative but insufficient to guarantee
better containment.

The Katz cases clarify why evaluating all four bases matters. Digg and
YouTube have identical HMI-Katz and raw-Katz selected sets and unchanged
coverage; Google+ has almost identical sets. In contrast, Betweenness
often changes the selected set substantially. A single HMI label would
hide these different responses to the same marginal-coverage rule.

\begin{table*}[t]
\centering
\small
\setlength{\tabcolsep}{4pt}
\caption{Minimum--maximum adaptive centrality weight across every primary selection step, for all 40 verified selector replays. Values are rounded to two decimals; these are deterministic ranges, not confidence intervals.}
\label{tab:compact_theta_ranges}
\begin{tabular}{lrrrr}
\toprule
Network & Degree & PageRank & Katz & Betweenness \\
\midrule
Twitter & $0.22$--$0.52$ & $0.07$--$0.33$ & $0.27$--$0.56$ & $0.10$--$0.41$ \\
Google+ & $0.35$--$0.55$ & $0.19$--$0.36$ & $0.49$--$0.69$ & $0.11$--$0.24$ \\
PGP & $0.33$--$0.53$ & $0.33$--$0.51$ & $0.32$--$0.61$ & $0.20$--$0.41$ \\
Facebook & $0.45$--$0.52$ & $0.49$--$0.56$ & $0.44$--$0.52$ & $0.27$--$0.38$ \\
Epinions & $0.43$--$0.63$ & $0.46$--$0.55$ & $0.46$--$0.60$ & $0.20$--$0.51$ \\
Digg & $0.30$--$0.51$ & $0.17$--$0.41$ & $0.80$--$0.89$ & $0.10$--$0.39$ \\
Weibo & $0.33$--$0.52$ & $0.31$--$0.50$ & $0.29$--$0.60$ & $0.17$--$0.44$ \\
Munmun Twitter & $0.25$--$0.50$ & $0.12$--$0.34$ & $0.25$--$0.50$ & $0.08$--$0.28$ \\
YouTube & $0.23$--$0.60$ & $0.22$--$0.52$ & $0.55$--$0.98$ & $0.15$--$0.49$ \\
Enron & $0.43$--$0.54$ & $0.37$--$0.54$ & $0.42$--$0.55$ & $0.15$--$0.31$ \\
\bottomrule
\end{tabular}
\end{table*}

\begin{figure*}[t]
\centering
\includegraphics[width=\textwidth]{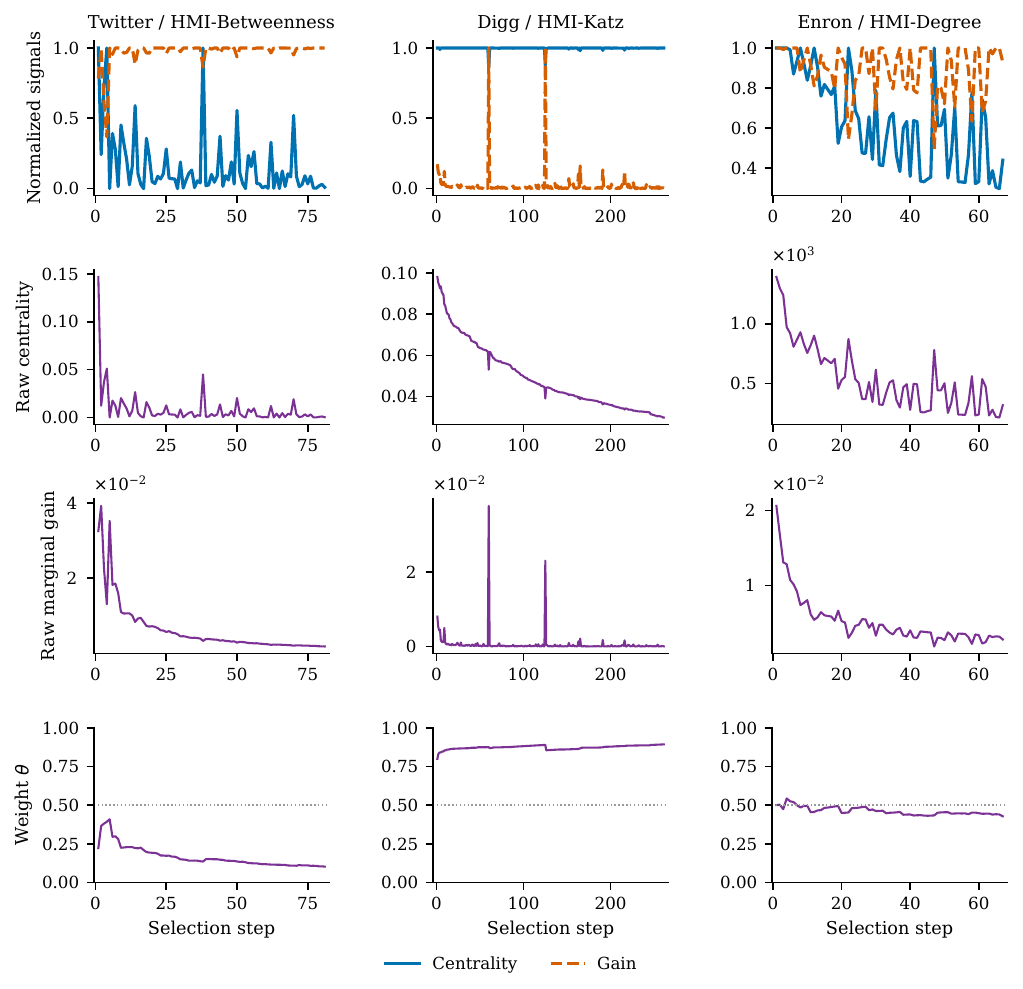}
\caption{Three contrasting selector regimes. Columns show Twitter
HMI-Betweenness, Digg HMI-Katz and Enron HMI-Degree. Rows retain normalized
selected-node signals, raw centrality, raw Jeffreys-marginal gain and the
adaptive centrality weight. The dotted line in the final row is $0.5$.
No smoothing or diffusion-seed averaging is applied. Cases illustrate
strong benefit, unchanged selection and an adverse primary mean; the
complete 40-selector ranges appear in Table~\ref{tab:compact_theta_ranges}.}
\label{fig:compact_selection}
\end{figure*}

Table~\ref{tab:compact_theta_ranges} retains the full ten-network view
of adaptive mixing in one table. Figure~\ref{fig:compact_selection}
examines three interpretable cases: Twitter HMI-Betweenness with a strong
matched benefit, Digg HMI-Katz with an unchanged selected set, and Enron
HMI-Degree with the primary adverse mean. The figure retains the four original
diagnostic quantities---normalized signals, raw centrality, raw gain and
$\theta$---alongside the complete cross-network range summary.

Each point describes the node chosen immediately before its addition.
Normalization uses all remaining candidates, and $\theta$ depends on the
dispersion across that full candidate set, not merely the two displayed
selected-node signals. Raw scales differ between centralities. Neither
raw selected centrality nor raw gain must decrease monotonically because
the score trades off both quantities. These deterministic traces describe
the ranking process; the following ablation evaluates the contribution of
adaptive mixing to diffusion outcomes.

\subsubsection{Ablation of the centrality--coverage mixing weight}
\label{app:theta_ablation}

The selection traces describe how the adaptive weight evolves, but do not
establish whether adapting it improves containment. We therefore compare
four conditions: the adaptive rule, a fixed $\theta=0.5$, pure centrality
($\theta=1$), and pure Jeffreys-marginal coverage gain ($\theta=0$).
The graph, budget, marginal-gain definition, candidate normalization,
and diffusion parameters are held fixed. Pure coverage retains the chosen
base only through the selector's existing deterministic candidate order
for breaking ties; it assigns no centrality contribution to the score.

\paragraph{Choice of base and evaluation scope.}
On each network, we use the adaptive HMI base with the lowest mean final
infections in the primary 100-seed experiment. Betweenness is used for
Twitter, Google+, Facebook, YouTube, and PGP; Katz for Weibo, Munmun Twitter,
and Enron; Degree for Digg; and PageRank for Epinions.
This choice was frozen before the ablation outcomes were examined.
The study is a retrospective component ablation conditional on the observed
primary winner. It neither evaluates all four bases under every weight nor
provides an independent test of a rule for choosing the best base.

All conditions use the primary profile: a 0.1\% debunker budget, 1\% initial
infections, $(\alpha,\beta,\gamma)=(0.2,0.15,0.1)$, an active spreading duration of two timesteps,
immediate intervention, and coverage neighborhoods containing each
candidate and its directly reachable nodes.
The 100 primary diffusion seeds are retained on every network.

\paragraph{Pairing and reuse.}
Every network--seed pair has the same exact initial infected nodes in all
four conditions. An original source list is retained whenever it is disjoint
from every ablation selection. Otherwise, only conflicting source nodes
are replaced, uniformly without replacement from eligible nodes outside
the original twelve-method debunker union, all four ablation selections,
and the retained sources. A separate seeded stream generates these
replacements before any ablation outcome is inspected. No seed is excluded.
Adaptive and pure-centrality outcomes are reused only when the source list
is unchanged; affected cases are simulated again with the common repaired
list. The pure-centrality selection exactly matches its saved raw-base
ranking, and the adaptive selection matches its verified primary replay.

The completed grid contains 4,000 evaluations: 2,056 new diffusion
simulations and 1,944 reused primary records. Source lists required repairs
for 28 of the 1,000 network--seed pairs. Consequently, adaptive and
centrality means can differ slightly from the primary table even though
their selected debunkers are unchanged. These counts are separate from the
12,000-run primary experiment and the parameter sweeps.

\begin{table*}[t]
\centering\small
\setlength{\tabcolsep}{3pt}
\caption{Theta ablation at the primary setting. Mean $\pm$ population standard deviation of final ever-infected percentage over 100 paired seeds; lower is better. The base is fixed to the lowest-mean adaptive HMI in the primary experiment. Centrality uses $\theta=1$; coverage uses $\theta=0$. Each column is a fixed condition, with no choice based on ablation outcomes.}
\label{tab:theta_outcomes}
\begin{tabular}{llrrrr}
\toprule
Network & Base & Adaptive & $\theta=.5$ & Centrality & Coverage \\
\midrule
Twitter & Betweenness & $6.619\pm0.382$ & $9.470\pm0.472$ & $10.104\pm0.527$ & $6.619\pm0.382$ \\
Google+ & Betweenness & $3.284\pm0.114$ & $9.722\pm0.300$ & $11.725\pm0.357$ & $3.281\pm0.119$ \\
PGP & Betweenness & $6.251\pm0.459$ & $6.415\pm0.411$ & $6.537\pm0.463$ & $6.275\pm0.463$ \\
Facebook & Betweenness & $9.940\pm0.663$ & $11.549\pm0.772$ & $14.934\pm1.042$ & $10.125\pm0.724$ \\
Epinions & PageRank & $1.806\pm0.092$ & $1.808\pm0.091$ & $1.844\pm0.092$ & $1.805\pm0.081$ \\
Digg & Degree & $1.557\pm0.031$ & $1.564\pm0.031$ & $1.668\pm0.045$ & $1.566\pm0.030$ \\
Weibo & Katz & $1.932\pm0.071$ & $1.946\pm0.073$ & $2.155\pm0.101$ & $1.922\pm0.068$ \\
Munmun Twitter & Katz & $1.047\pm0.003$ & $1.055\pm0.003$ & $1.061\pm0.003$ & $1.047\pm0.003$ \\
YouTube & Betweenness & $1.586\pm0.017$ & $1.678\pm0.021$ & $1.711\pm0.020$ & $1.585\pm0.017$ \\
Enron & Katz & $1.494\pm0.129$ & $1.477\pm0.139$ & $1.534\pm0.183$ & $1.520\pm0.131$ \\
\bottomrule
\end{tabular}
\end{table*}

\paragraph{Outcome comparison.}
Adaptive mixing has lower mean final infections than equal weighting on
nine networks and has a higher mean on Enron.
Against pure centrality it has ten lower means.
These counts describe differences in means, rather than counting
statistically significant improvements. Across all 30 adaptive-versus-arm
contrasts jointly, Holm-adjusted tests retain seven reductions against
equal weighting and nine against pure centrality; no increase passes.
This adjustment does not remove the primary-base selection effect, so the
ablation remains exploratory. Against pure coverage the outcome
is more balanced: five lower means, one exact tie, and four higher means.
Table~\ref{tab:theta_coverage_effects} reports their magnitudes, rather than
treating all wins and losses as equally consequential.

\begin{table*}[t]
\centering\small
\setlength{\tabcolsep}{3pt}
\caption{Adaptive minus pure-coverage final infections. Negative values favor adaptive mixing. Percentage-point (pp) differences use graph population; relative change divides the mean difference by the pure-coverage mean. Intervals are pointwise 95\% paired $t$ intervals. The final column adjusts tests over all 30 component contrasts with Holm's procedure; neither intervals nor tests correct for primary-base selection.}
\label{tab:theta_coverage_effects}
\begin{tabular}{lrrrr}
\toprule
Network & Difference (pp) & Relative change (\%) & Paired 95\% interval (pp) & $p_{\mathrm H}$ \\
\midrule
Twitter & +0.00000 & +0.000 & $[+0.00000,\,+0.00000]$ & $1.0000$ \\
Google+ & +0.00284 & +0.087 & $[-0.01489,\,+0.02057]$ & $1.0000$ \\
PGP & -0.02407 & -0.384 & $[-0.11307,\,+0.06493]$ & $1.0000$ \\
Facebook & -0.18458 & -1.823 & $[-0.34278,\,-0.02639]$ & $0.2040$ \\
Epinions & +0.00088 & +0.049 & $[-0.01352,\,+0.01529]$ & $1.0000$ \\
Digg & -0.00938 & -0.599 & $[-0.01464,\,-0.00413]$ & $0.0091$ \\
Weibo & +0.00962 & +0.501 & $[+0.00201,\,+0.01724]$ & $0.1515$ \\
Munmun Twitter & -0.00029 & -0.028 & $[-0.00101,\,+0.00043]$ & $1.0000$ \\
YouTube & +0.00105 & +0.066 & $[-0.00199,\,+0.00408]$ & $1.0000$ \\
Enron & -0.02629 & -1.730 & $[-0.04680,\,-0.00578]$ & $0.1502$ \\
\bottomrule
\end{tabular}
\end{table*}

The largest relative improvement over pure coverage is on Facebook:
$-0.18458$ percentage points, or $1.823\%$ fewer infections relative to
pure coverage, corresponding to 81.13 fewer infected nodes per simulation
on average. Enron improves by $1.730\%$ relatively. The largest relative
loss is on Weibo: $+0.00962$ percentage points, or $0.501\%$ more,
corresponding to 8.52 additional infected nodes on average.
Ordinary paired 95\% intervals exclude zero in favor of adaptive mixing
on Facebook, Digg, and Enron, and in favor of pure coverage on Weibo.
The other nonzero differences have intervals crossing zero. After Holm correction across all 30 component contrasts, only
the Digg reduction passes ($p_{\mathrm H}=0.0091$): $-0.00938$ percentage
points, or $0.599\%$ relative to pure coverage. The Facebook and
Enron reductions and the Weibo increase do not pass this corrected test
(Table~\ref{tab:theta_coverage_effects}). The pointwise intervals remain
unadjusted, and even the corrected tests do not account for choosing the
primary-winning base. The evidence therefore does not establish a general
advantage of adaptive mixing over pure coverage, nor does failure to detect
a difference establish equivalence. Pure coverage remains a strong
comparator, with small, network-dependent observed differences.

\paragraph{Do the two conditions select the same nodes?}
\label{app:theta_overlap}
Let $A$ and $C$ denote the adaptive and pure-coverage selected sets, each
of size $k$. We report both the shared fraction $|A\cap C|/k$ and Jaccard
overlap $|A\cap C|/|A\cup C|$ in Table~\ref{tab:theta_overlap}.
These are exact comparisons of one deterministic selected set per
condition and network; they are not averages over diffusion seeds.

\begin{table*}[t]
\centering\small
\setlength{\tabcolsep}{3pt}
\caption{Exact overlap between adaptive and pure-coverage selected sets. Each method selects $k$ nodes; Different counts nodes unique to each set. Shared percentage is $100|A\cap C|/k$ and Jaccard percentage is $100|A\cap C|/|A\cup C|$. Each network has one deterministic set per condition, reused across diffusion seeds.}
\label{tab:theta_overlap}
\begin{tabular}{lrrrrr}
\toprule
Network & $k$ & Shared & Different & Shared (\%) & Jaccard (\%) \\
\midrule
Twitter & 81 & 81 & 0 & 100.00 & 100.00 \\
Google+ & 201 & 200 & 1 & 99.50 & 99.01 \\
PGP & 39 & 37 & 2 & 94.87 & 90.24 \\
Facebook & 43 & 37 & 6 & 86.05 & 75.51 \\
Epinions & 75 & 62 & 13 & 82.67 & 70.45 \\
Digg & 261 & 246 & 15 & 94.25 & 89.13 \\
Weibo & 88 & 86 & 2 & 97.73 & 95.56 \\
Munmun Twitter & 465 & 351 & 114 & 75.48 & 60.62 \\
YouTube & 1134 & 1121 & 13 & 98.85 & 97.73 \\
Enron & 67 & 60 & 7 & 89.55 & 81.08 \\
\bottomrule
\end{tabular}
\end{table*}

The mean shared fraction is $91.90\%$ with equal weight per network;
the corresponding mean Jaccard overlap is $85.93\%$.
Pooling within-network selected-node counts gives 2,281 shared selections
out of 2,454, or $92.95\%$; this places more weight on larger budgets.
Node identifiers are never compared across different graphs.
Twitter selects exactly the same set, although the selection order differs,
which explains its identical diffusion outcomes under simultaneous
intervention. Google+ differs by only one selected node per method,
and YouTube shares 1,121 of 1,134 nodes.

The lowest shared fraction is on Munmun Twitter: 351 of 465 nodes
($75.48\%$), leaving 114 different nodes per method. Nevertheless,
its final infection difference is only $-0.00029$ percentage points
($-0.028\%$ relatively). High set overlap therefore helps explain similar
outcomes on several graphs, while different sets can also produce similar
containment. These observations do not establish that centrality is
universally redundant or that the nonshared nodes have identical roles.

\subsection{Accompanying evidence and reproducibility}
\label{app:result_archive}
\label{app:restored_artifacts}
\label{app:dataset_provenance}
\label{app:latest_reproducibility}

The evidence archive retains the complete outcomes and diagnostic arrays
behind the condensed tables and illustrative figures. Primary, sweep, theta-ablation, and selector-replay results have separate provenance; their sample counts must not be pooled. The primary campaign
contains 12,000 records. The sweeps contain 104,000 raw records, including
1,600 Google+ delay records imported once from an earlier completed block;
reused canonical primary points are not additional simulations. The
combined sweep CSV contains 1,280 method--setting rows. The theta ablation
contains 4,000 evaluations, of which 2,056 are new and 1,944 are reused
primary outcomes. 
The dataset manifests record original URLs, archive and graph hashes,
direction conventions and preprocessing. Followers, trust, certification,
wall posts and email are heterogeneous relations standardized to one
exposure model; Enron is reversed to recipient-to-sender. Edge weights,
timestamps and repeated interactions are discarded. The study therefore
tests robustness across graph structures, rather than estimating causal
platform-specific intervention effects.

The primary generator validates completed blocks, artifact hashes and
sizes, exact method/seed grids, common initial sources, source/debunker
separation, shared parameters and raw-to-aggregate agreement. Saved
trajectory checks additionally verify state conservation and secondary
metrics. Sweep checks verify the corresponding setting grids and pairing;
ablation checks preserve repaired-source provenance. The seed sampler uses
seed 20260730 to choose the following 100 distinct integers from $[0,10^6)$,
reused in their saved order on every network and for every method:

\begin{quote}
\footnotesize
733980, 626956, 183324, 266579, 559726, 176907, 35514, 769350, 589887, 418549,
589641, 621567, 62242, 180631, 565512, 346097, 597739, 620959, 525138, 490701,
856309, 373162, 74565, 802825, 543356, 259262, 793414, 309308, 994419, 377081,
206214, 237404, 422941, 740831, 292942, 256405, 72256, 728140, 68498, 500232,
400276, 412115, 241326, 437813, 868494, 851602, 414266, 708616, 915235, 745126,
509910, 561971, 692144, 466768, 46227, 797062, 364114, 103092, 733783, 129832,
990793, 132824, 282838, 256103, 270625, 369971, 612527, 513577, 128397, 193901,
285594, 2255, 371672, 630058, 665513, 757352, 705989, 662203, 75648, 508738,
387318, 830512, 864250, 696213, 736515, 454365, 87749, 371699, 328795, 763341,
481952, 405103, 867583, 349407, 235344, 858378, 606749, 651259, 554763, 956135
\end{quote}

The accompanying evidence index records the complete numerical summaries,
input hashes and immutable raw-campaign paths.

\section{Experimental Details}
\label{app:hmi_experimental_details}

\subsection{Network Sources and Preprocessing}
The campaign contains ten directed social, trust, interaction, and
communication graphs. Twitter comes from the SNAP ego-network collection, and Epinions is the SNAP Web of Trust graph. Google+ is the \texttt{soc-google-plus}
graph distributed by Network Repository.
PGP is the 2009 Web of Trust graph distributed through Netzschleuder. Facebook Wall Posts and
YouTube Social Links are distributed through KONECT.
Digg Friends comes from the Digg 2009 archive; Weibo
Followers comes from the public emotion-contagion data release. Munmun Twitter is the follower network from
De Choudhury et al., distributed through KONECT; Enron is the directed email
graph distributed by Network Repository.

The campaign uses cached directed, unweighted graphs with repeated directed
pairs collapsed, the largest weakly connected component retained, and node
identifiers relabeled contiguously. The exact cached Twitter and Google+
graphs retain 14 and 1,713 self-loops, respectively; the other eight graphs
have none. These arcs remain in the campaign graph and its coverage
diagnostics. Only the simple undirected projection used for triangle and
clustering statistics removes self-loops. Stored directions preserve each source relation, except Enron,
which is reversed from sender--recipient to recipient--sender so that the
standardized exposure convention remains consistent. PGP metadata, Facebook timestamps, and repeated interaction counts
are not used. Table~\ref{main-tab:dataset_statistics_compact} of the main text reports the exact
processed sizes. The collection ranges from 39,796 to 1,134,885 nodes and
from 1.80 to 92.34 directed edges per node.

\subsection{Paired Statistical Analysis}
We additionally analyze the saved per-seed endpoints using two-sided paired
$t$ tests of zero mean difference over the 100 seeds (99 degrees of freedom).
Source-set hashes are checked before pairing, including repaired sources
in the theta ablation. For these tests, differences are HMI minus its
comparator, so negative values favor HMI. We apply Holm's step-down
correction
at the 5\% level separately to three families: all 40 fixed HMI--base
final-infection comparisons; all 80 matched peak-infection and AUC
comparisons jointly; and all 30 adaptive-versus-component ablation
comparisons jointly. We denote the resulting adjusted value by $p_{\mathrm H}$.
The reported 95\% intervals remain pointwise, not simultaneous intervals;
an interval excluding zero need not pass the corrected test.

This is a post hoc analysis of the existing runs, not a new experiment.
Inference concerns variation across diffusion seeds on the fixed graphs
and settings, rather than a random sample of networks or real interventions.
Holm correction addresses multiplicity within each family, not selection
of the primary-winning base for the ablation; those component comparisons
remain exploratory. Best-of-family summaries and sweep win counts remain
descriptive. Failure to reject a difference does not establish equivalence,
and statistical significance does not by itself imply a practically large
effect. Effect sizes and intervals accompany the tests in the numerical
archive.

\subsection{Comparison Methods}
\label{app:comparison_methods}
The baselines represent complementary philosophies of influential-node
selection used in network science and information diffusion: (i) local
structure-based methods, which prioritize strong immediate neighborhoods;
(ii) global influence-based methods, which favor nodes embedded in globally
central regions; (iii) path- and bottleneck-based methods, which target bridges
in information flow; and (iv) dispersion-oriented methods, which discourage
redundant selections.  This comparison tests whether HMI benefits from
balancing structural influence with marginal coverage rather than from
specializing in only one of these classes.
Below, $n=|V|$ denotes the number of nodes.

\paragraph{Degree (local structure-based).}
In our directed setting, Degree uses in-degree, defined as the number of
nodes directly reachable from a node, $k_i=|F(i)|$. The $K$ nodes with the largest
in-degree are selected, reflecting the intuition that nodes with large direct audiences can disseminate corrective information widely
\cite{freeman1978centrality}.

\paragraph{Leverage centrality (local structure-based).}
Leverage centrality measures how much more or less connected a node is than
its neighbors.  Its undirected form is
\begin{equation}
    LC_i=\frac{1}{k_i}\sum_{j\in N(i)}\frac{k_i-k_j}{k_i+k_j}.
\end{equation}
We use the directed adaptation from the implementation: $k_i$ is the
in-degree of node $i$, $N(i)=F(i)$ is its directly reachable set, and each directly reachable node is
compared using its own in-degree. The implementation skips zero-in-degree neighbors while retaining the total neighbor count in the denominator, and sets the score to zero for a node with no directly reachable nodes. A high value therefore identifies a node
that is locally dominant among its directly reachable neighbors
\cite{joyce2010leverage}.

\paragraph{PageRank (global influence-based).}
PageRank models a random walk on the directed graph and assigns higher scores
to nodes receiving links from already influential nodes
\cite{page1999pagerank}.  We use NetworkX \cite{hagberg2008networkx} PageRank on the standardized graph
with the conventional damping factor of $0.85$; the top $K$ scores form the debunker set.
The score vector satisfies
\[
    \mathbf r=dP^{\mathsf T}\mathbf r+\frac{1-d}{n}\mathbf 1,
    \qquad d=0.85,
\]
where $\mathbf 1$ is the all-ones vector and $P$ chooses uniformly among
each node's information sources. For nodes with no such sources, the corresponding row
of $P$ is uniform over all $n$ nodes.

\paragraph{LeaderRank-inspired baseline (global influence).}
LeaderRank augments the directed graph with a ground node connected
bidirectionally to every original node \cite{lu2011leaderrank}.  We apply
PageRank with damping factor $0.85$ to this augmented graph, remove the ground
node, renormalize the remaining scores, and select the top $K$ nodes.
Writing $\mathbf r^+$ for PageRank on the augmented graph and $g$ for its
ground node, the score of an original node is
\[
    \ell_i=\frac{r_i^+}{\sum_{j\in V}r_j^+}
          =\frac{r_i^+}{1-r_g^+},\qquad i\in V.
\]
This is a ground-node PageRank adaptation: standard LeaderRank uses an undamped walk and redistributes the ground-node score. Tables retain the campaign label ``LeaderRank''; its reported results refer to this adaptation.

\paragraph{Katz centrality (global influence-based).}
Katz centrality extends degree by incorporating indirect influence, with the
contribution of longer paths decreasing geometrically
\cite{katz1953status}.  We use the directed graph with attenuation
$\alpha_{\mathrm{Katz}}=0.005$ and exogenous weight $1.0$, matching the current
implementation. The unnormalized scores $x_i$ and normalized scores $c_i$ satisfy
\[
    x_i=1+\alpha_{\mathrm{Katz}}\sum_{j\in F(i)}x_j,
    \qquad c_i=\frac{x_i}{\sqrt{\sum_{j\in V}x_j^2}}.
\]
The usual nonnegative Katz path-sum interpretation requires $\alpha_{\mathrm{Katz}}\rho(A)<1$, where $\rho(A)$ is the spectral radius of the adjacency matrix $A$, with $A_{ij}=1$ when $(i,j)\in E$ and zero otherwise. Scores are computed by sparse power iteration with at most 1,000 iterations, an $L_1$ change tolerance of $|V|10^{-6}$, and final Euclidean normalization; failure to converge raises an error.

\paragraph{Betweenness centrality (path-based).}
Directed betweenness measures how frequently a node lies on shortest paths
between ordered node pairs. Its normalized score is
\[
    B(v)=\frac{1}{(n-1)(n-2)}
    \sum_{\substack{s,t\in V\setminus\{v\}\\s\ne t}}
    \frac{\sigma_{st}(v)}{\sigma_{st}},
\]
where $\sigma_{st}$ is the number of shortest directed paths from $s$ to
$t$ and $\sigma_{st}(v)$ counts those passing through $v$. Unreachable
pairs contribute zero. High-betweenness nodes act as bridges or
bottlenecks in information flow and are therefore natural intervention points
\cite{freeman1977betweenness}.  For scalability, both the unmodified baseline
and the signal supplied to HMI use the same sampled-pivot approximation
\cite{brandes2007centrality} with a fixed random seed:
normalized, unweighted, non-endpoint betweenness on the reversed graph, using
$200$ sampled pivots and random seed $0$.

\paragraph{VoteRank (dispersion-oriented).}
VoteRank iteratively identifies multiple spreaders while reducing redundancy.
In the directed adaptation, an unselected candidate receives votes from
its directly reachable nodes. If $a_j$ is the voting ability of reachable node $j$, the candidate's score is
\[
    s_i=\sum_{j\in F(i)}a_j,\qquad a_j=1\ \text{initially}.
\]
After selecting a candidate $i^\star$ with the largest score, set
$a_{i^\star}=0$ and update $a_j\leftarrow\max\{0,a_j-1/\bar{k}\}$ for
$j\in F(i^\star)$, where $\bar{k}=|E|/n$ is the mean number of directly reachable nodes.
Selected nodes are excluded from subsequent choices. This reduces repeated
selection from the same local region \cite{zhang2016voterank}.
The implementation uses the reversed graph and an incremental VoteRank routine, falling back to NetworkX for graphs with self-loops. It requests up to $\lceil0.02|V|\rceil$ ranked nodes, stopping earlier if no positive score remains, and uses the leading $K$ cached scores for the primary $0.1\%$ budget.

\paragraph{Collective influence (dispersion-oriented).}
The collective influence score of node $i$ at radius $l$ is
\begin{equation}
    CI_l(i)=(k_i-1)\sum_{j\in\partial Ball(i,l)}(k_j-1),
\end{equation}
where $\partial Ball(i,l)$ contains nodes exactly $l$ hops from $i$
\cite{morone2015influence}.  We use $l=2$, in-degree for $k$, and traverse
from each node to its directly reachable nodes at each hop. Scores are set to zero when $k_i\leq1$. We select the top $K$ scores once, without graph removal or reranking: this is a static directed adaptation, rather than the adaptive dismantling algorithm of the original work.

\paragraph{Four fixed HMI variants.}
The four proposed methods are HMI-Degree, HMI-PageRank, HMI-Katz, and
HMI-Betweenness. They share the Jeffreys-marginal response mode, coverage neighborhoods
containing each candidate and its directly reachable nodes, dynamic min--max normalization,
and adaptive mixture weight. They differ only in the base-centrality signal. No HMI
variant uses $\alpha$, $\beta$, $\gamma$, or an evaluation outcome to select
nodes. All four bases are evaluated directly; no held-out base-centrality selection is performed.

\bibliography{references,hmi_centrality_refs}